\documentclass[draftclsnofoot, onecolumn, 12pt]{IEEEtran}

\usepackage{mathtools}

\usepackage{floatflt}

\usepackage{color}
\usepackage{cite}

\usepackage{graphicx}
\usepackage{amsmath}
\usepackage{amsthm}
\usepackage{amssymb}

\usepackage{verbatim}

\usepackage{psfrag,array}

\usepackage{subfigure}

\usepackage{stfloats}

\usepackage{todonotes}

\usepackage{amsmath}
\usepackage{epstopdf}
\usepackage{algorithmic}

\newcommand{\p}[1]{\mathop{\mbox{\it p} } }

\renewcommand{\vec}[1]{\ensuremath{\boldsymbol{#1}}}
\newcommand{\be}{\begin{equation}}
\newcommand{\ee}{\end{equation}}
\newcommand{\ba}{\begin{array}}
\newcommand{\ea}{\end{array}}
\newcommand{\bea}{\begin{eqnarray}}
\newcommand{\eea}{\end{eqnarray}}
\newcommand{\bean}{\begin{eqnarray*}}
\newcommand{\eean}{\end{eqnarray*}}

\newcommand{\rmh}{^{\dag}}
\newcommand{\rmt}{^{\rm T}}

\definecolor{white}{rgb}{1,1,1}

\newtheorem{property}{Property}

\newtheorem{corollary}{Corollary}

\begin{document}

\title{Beyond Massive-MIMO: The Potential of Data-Transmission with Large Intelligent Surfaces}
\author
{
Sha Hu, Fredrik Rusek, and Ove Edfors
\thanks{The authors are with the Department of Electrical and Information Technology, Lund University, Lund, Sweden $\left(\text{\{firstname.lastname\}@eit.lth.se}\right)$.}
\thanks{This paper has been presented in part in IEEE Vehicular Technology Conference (VTC), Sydney, Australia, 4-7 Jun. 2017. }
}
\maketitle

\begin{abstract}
In this paper, we consider the potential of data-transmission in a system with a massive number of radiating and sensing elements, thought of as a contiguous surface of electromagnetically active material. We refer to this as a large intelligent surface (LIS). The \lq\lq{}LIS\rq\rq{} is a newly proposed concept, which conceptually goes beyond contemporary massive MIMO technology, that arises from our vision of a future where man-made structures are electronically active with integrated electronics and wireless communication making the entire environment \lq\lq{}intelligent\rq\rq{}. 

We firstly consider capacities of single-antenna autonomous terminals communicating to the LIS where the entire surface is used as a receiving antenna array. Under the condition that the surface-area is sufficiently large, the received signal after a matched-filtering (MF) operation can be closely approximated by a sinc-function-like intersymbol interference (ISI) channel. Secondly, we analyze the capacity per square meter (m$^2$) deployed surface, $\hat{\mathcal{C}}$, that is achievable for a fixed transmit power per volume-unit, $\hat{P}$; the volume-unit can be m, m$^2$, and m$^3$ depending on the scenario under investigation. As terminal-density increases, the limit of $\hat{\mathcal{C}}$ achieved when the wavelength $\lambda$ approaches zero is $\hat{P}/(2N_0)$ [nats/s/Hz/volume-unit], where $N_0$ is the spatial power spectral density (PSD) of the additive white Gaussian noise (AWGN). Moreover, we also show that the number of independent signal dimensions per m deployed surface is $2/\lambda$ for one-dimensional terminal-deployment, and $\pi/\lambda^2$ per m$^2$ for two and three dimensional terminal-deployments. Thirdly, we consider implementations of the LIS in the form of a grid of conventional antenna elements and show that, the sampling lattice that minimizes the surface-area of the LIS and simultaneously obtains one signal space dimension for every spent antenna is the hexagonal lattice. Lastly, we extensively discuss the design of the state-of-the-art low-complexity channel shortening (CS) demodulator for data-transmission with the LIS.
\end{abstract}


\section{Introduction}

We envision a future where man-made structures become more and more electronically active, with integrated electronics and wireless communication making the entire environment intelligent. A Large Intelligent Surface (LIS) is an entirely new concept in wireless communication \cite{HRE171, HRE172}, and makes new and disruptive applications which require high energy efficiency and transmission reliability, low latency and ability to interact with the environment, possible. LISs allow for an unprecedented focusing of energy in three-dimensional space which enables wireless charging, remote sensing with extreme precision and unprecedented data-transmissions. This makes it possible to fulfill the most grand visions for the next generation of communication systems and the concept of Internet of Things \cite{IoT, AZ14}, where billions of devices are expected to be connected to the Internet.

LIS can be seen as an extension of massive MIMO\cite{M10, MM12, MM14}, but it scales up beyond the traditional antenna array concept. In Fig. {fig11}, we show an example of three terminals communicating to a LIS in indoor and outdoor scenarios, respectively. A concept somewhat similar to what we call LIS seems to be first mentioned in the eWallpaper project at UC Berkeley \cite{ewall}, where the ultimate vision is to fabricate wall papers that are electromagnetically active and has built-in processing power. However, no analysis has been carried out on information-transfer capabilities. Rather, the efforts have been directed towards hardware and fabrication aspects of intelligent surfaces.

In this paper, we take a first look at the information-transfer capabilities of the LIS in the uplink (UL). For analytical tractability we assume an ideal situation where no scatterers or reflections are present, yielding a perfect line-of-sight (LoS) propagation scenario, and each autonomous terminal is assumed to propagate an isotropic signal. We show that, the limit of $\hat{\mathcal{C}}$ which is the normalized capacity per volume-unit in space, achieved when wavelength $\lambda$ approaches zero, is $\frac{\hat{P}}{2N_0}$ [nats/s/Hz/volume-unit], where $\hat{P}$ is the transmit power per volume-unit and $N_0$ is the spatial power spectral density (PSD) of additive white Gaussian noise (AWGN). In particular, we show that for an infinitely large LIS, $\frac{2}{\lambda}$ terminals can be spatially multiplexed per m deployed surface\footnote{We here assume a rectangular LIS and measure its size only by its length, while its height is assumed to extend infinitely.} for one-dimensional terminal-deployment, and $\frac{\pi}{\lambda^2}$ terminals can be spatially multiplexed per m$^2$ deployed surface-area for two and three dimensional terminal-deployments, respectively. We also demonstrate through numerical simulations that, with a fairly small LIS deployed in a medium sized room, around 100 terminals can be accommodated in the UL with only a minuscule per-terminal capacity loss compared to a case where only one terminal is present, due to effective interference suppressing with the LIS. 

Then, we also analyze optimal implementation of the LIS based on sampling theory \cite{PM62, KA05}, and show that, the hexagonal lattice minimizes the surface-area of the LIS while simultaneously obtaining one independent signal dimension for every spent antenna-element on the LIS. With the same number of independent signal dimensions achieved, the hexagonal lattice yields 23\% surface-area saving over a rectangular lattice. Lastly, we extensively discuss the design of low-complexity demodulation for data-transmission in the UL. A particularly well suited method is channel shortening (CS) \cite{RP12, HR17j1,HR161,HR162} which we investigate in detail. CS can provide a complexity-performance trade-off between linear minimum-mean-square-error (LMMSE) \cite{MS02} and the optimal BCJR \cite{BCJR} demodulator via selecting different intersymbol interference (ISI) depth $\nu$. We show through simulations that, the LMMSE demodulator (which is a special case of the CS demodulator with $\nu\!=\!0$) performs close to the optimal receiver, when the deployed terminal-density is close to the number of independent signal dimensions that can be achieved. 

\begin{figure}[t]
\begin{center}
\vspace*{-0mm}
\hspace*{-2mm}
\scalebox{0.65}{\includegraphics{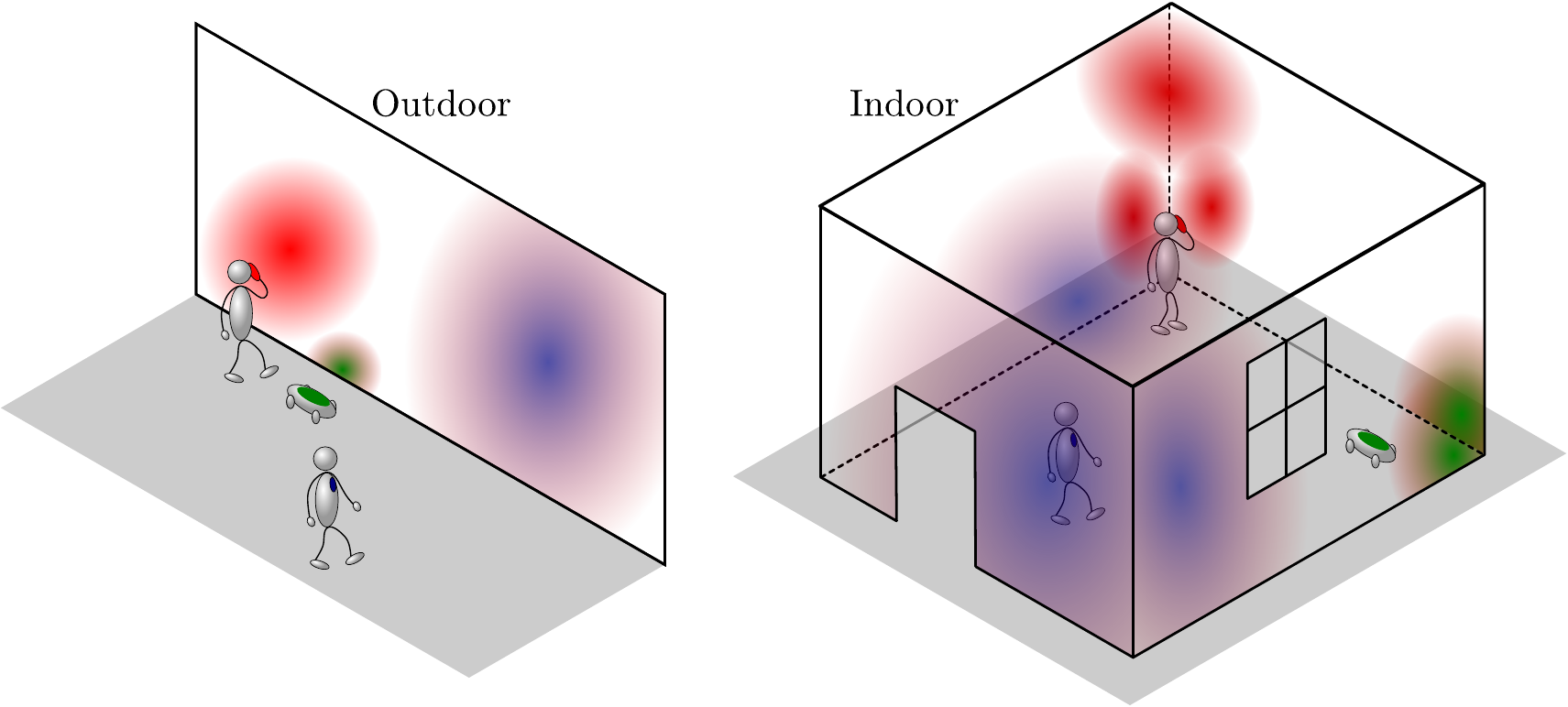}}
\vspace*{-8mm}
\caption{\label{fig1}An example of three terminals communicating to a LIS in indoor and outdoor scenarios, respectively.}
\vspace*{-12mm}
\end{center}
\end{figure}

The rest of the paper is organized as follows. In Sec. II we describe the received signal model for LIS and introduce a sinc-function based approximation of the ISI channel after a match-filter (MF) procedure for analytical tractability. In Sec. III we analyze the capacities for both the optimal and MF receivers for one, two and three dimensional terminal-deployments, and put a special interest on the independent signal dimensions that can be harvest per m$^2$ deployed surface. In Sec. IV we derive the optimal lattice that minimizes the surface-area of a LIS while achieving one independent signal dimension for every spent antenna. In Sec. V we elaborate a low-complexity demodulator design based on CS. Numerical results are presented in Sec. VI and Sec. VII summarizes the paper.

\subsubsection*{Notation}
Throughout this paper, superscripts $(\cdot)^{-1}$, $(\cdot)^{\frac{1}{2}}$, $(\cdot)^{\ast}$,
$(\cdot)\rmt$ and $(\cdot)\rmh$ stand for the inverse, matrix square root, complex conjugate, transpose, and
Hermitian transpose, respectively. Boldface letters indicate vectors and boldface uppercase letters designate matrices. We also reserve $a_{m,n}$ to denote the element at the $m$th row and $n$th column of matrix $\vec{A}$, $a_{m}$ to denote the $m$th element of vector $\vec{a}$, and $\vec{I}$ to represent the identity matrix. The operators \lq{}$\mathcal{R}\{\cdot\}$\rq{} and \lq{}$\mathrm{Tr}(\cdot)$\rq{} take the real part and the trace of the arguments, \lq{}$\star$\rq{} denotes linear convolution, and \lq{}$\mathbb{E}[\cdot]$\rq{} is the expectation operation.

\section{Received Signal Model at LIS for Multiple Terminals}

\subsection{Narrow-band Received Signal Model at the LIS}

\begin{figure}[t]
\begin{center}
\vspace*{-2mm}
\hspace*{-8mm}
\scalebox{0.29}{\includegraphics{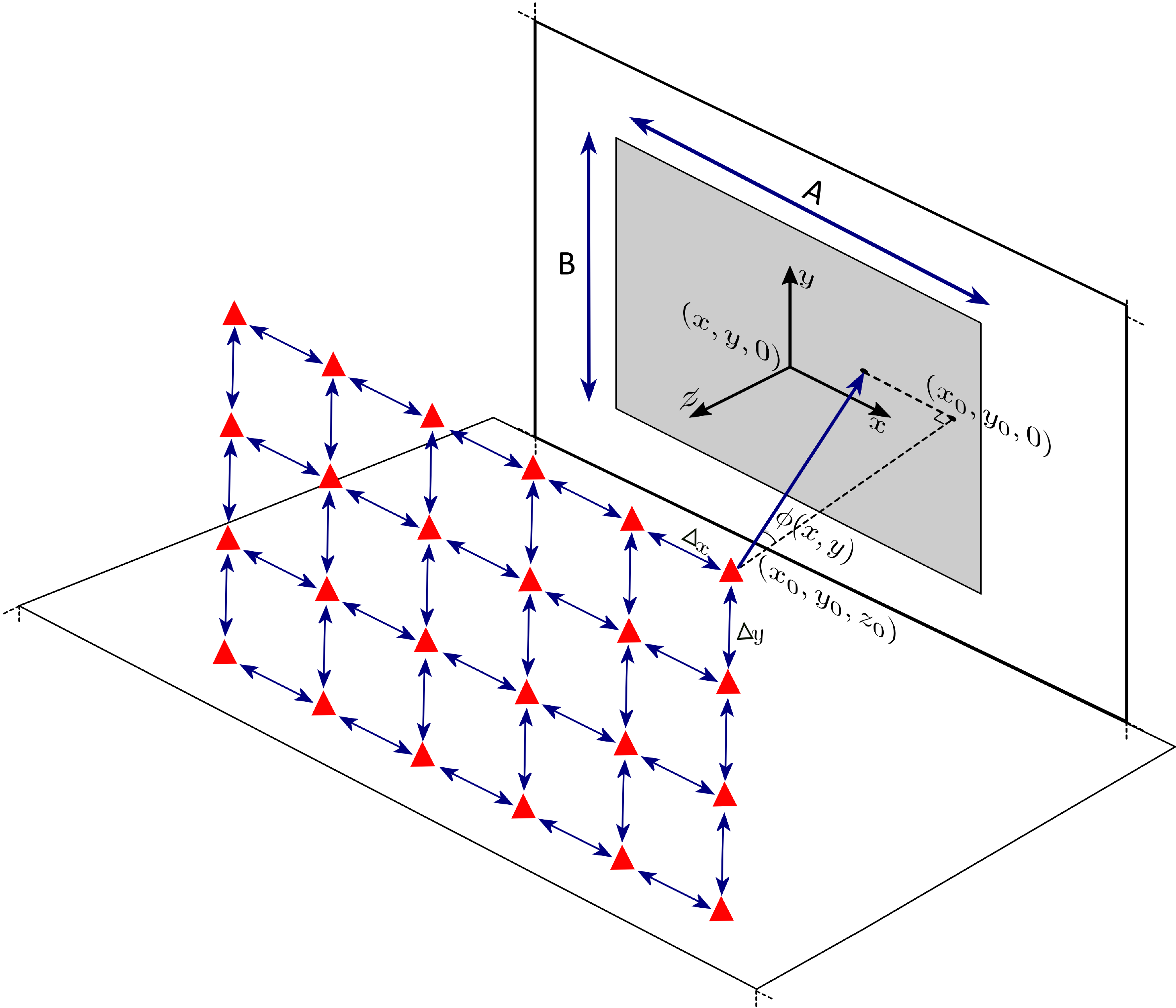}}
\vspace*{-7mm}
\caption{\label{fig2}The radiating model of transmitting signal to the LIS with terminals deployed in front of the LIS. We integrate the received signal of each point-element over the whole surface-area spanned by the LIS. Hence, for each point-element on the LIS, the Fraunhofer distance \cite{F10} is infinitely small, and the received signal model (\ref{md1}) holds for both near-field and far-field scenarios with respect to the LIS.}
\vspace*{-12mm}
\end{center}
\end{figure}

We consider the transmission from $K$ autonomous single-antenna terminals located in a three-dimensional space to a two-dimensional LIS deployed on a plane as shown in Fig. \ref{fig2}. Expressed in Cartesian coordinates, the LIS center is located at $x\!=\!y\!=\!z\!=\!0$, while terminals are located at $z\!>\!0$ and arbitrary $x$, $y$ coordinates. For analytical tractability, we assume a perfect LoS propagation. The $k$th terminal located at $(x_k,y_k,z_k)$ transmits data symbols $u_k[m]$ with power $P_k$ (per Hz), and $u_k[m]$ are assumed to be independent Gaussian variables with zero-mean and unit-variance. We denote $\lambda$ as the wavelength and $T$ as the symbol period, and consider a narrow-band system where the transmit times from terminals to the LIS are negligible compared to $T$ which results in no temporal interference. The received baseband signal at the LIS location $(x,y,0)$ corresponding to the $k$th terminal at time $t$ is
\bea \label{md} \tilde{s}_{x_k,\,y_k,\,z_k}(x,y,t)=s_{x_k,\,y_k,\,z_k}(x,y)\sqrt{P_k}\sum\limits_{m=-\infty}^{\infty}u_k[m]\mathrm{sinc}_T\left(t-mT\right)\!+n(x,y,t), \eea
where \lq{}sinc$_T$(t)\rq{} is a unit-energy sinc pulse with two-sided bandwidth $W\!=\!1/T$, and the noise-term $n(x,y,t)$ is independent over positions $(x,y)$, and modeled as wide-sense stationary (WSS) Gaussian process with zero-mean and a PSD $N_0$ at each position ($x,y$) on the LIS. 

The effective channel $s_{x_k,\,y_k,\,z_k}(x,y)$ can be modeled as
\bea  \label{nbs1} s_{x_k,\,y_k,\,z_k}(x,y)=\sqrt{\varepsilon_{L}\cos\phi(x,y)}\exp\!\left(\!-2\pi j f_c \Delta_k(x,y)\right) \!,
 \eea
where the parameters are defined as follows: $\varepsilon_L\!=\!1/(4\pi\eta_k)$ is the free-space path-loss; $\phi(x,y)$ is the arrival-of-angle (AoA) and $\cos\phi(x,y)\!=\!z_k/\sqrt{\eta}$ (see Fig.2); $f_c$ is the carrier-frequency; the transmit time from the $k$th terminal to the location $(x,y,0)$ is $\Delta_k(x,y)\!=\!\sqrt{\eta_k}/c$, with $c$ being the speed-of-light and the metric
\bea \eta_k=z_k^2+(y-y_k)^2+(x-x_k)^2.\eea 
Inserting them back into (\ref{nbs1}) yields the effective channel of the LIS for the $k$the terminal as
\bea \label{md1} s_{x_k,\,y_k,\,z_k}(x,y)=\frac{\sqrt{ z_k}}{2\sqrt{\pi}\eta_k^{3/4}}\exp\!\left(\!-\frac{2\pi j\sqrt{\eta_k}}{\lambda}\right)\!.   \eea

Signal model (\ref{md1}), which has also been discussed in \cite[Proposition 1]{HRE172}, is more accurate than what is usually considered in traditional large antenna-array systems \cite{SW94}, where in the latter case terminals are assumed to be in the far-field and a planar-wave approximation is used in (\ref{nbs1}) and the term $\cos\phi(x,y)$ is approximated by 1.

\subsection{Received Signal Model for Multiple Terminals with MF Procedure}
Based on (\ref{md}), the received signal at the LIS location $(x, y, 0)$ comprising signals from all $K$ terminals equals
\bea \label{rxyt} r(x,y,t) = \sum_{k=0}^{K-1}\sum\limits_{m=-\infty}^{\infty} s_{x_k,y_k,z_k}(x,y) \sqrt{P_k}u_k[m]\mathrm{sinc}_T\left(t-mT\right) +n(x,y,t).\eea
Given received signal (\ref{rxyt}) across the LIS, optimum processing includes applying both a spatial and a temporal correlator to each transmit signal, a procedure we call \lq{}MF\rq{}. The discrete received signal at sampling time $mT$ after the MF process is
{\setlength\arraycolsep{2pt} \bea \label{rkm} r_k[m] &=& \sqrt{P_k}\iint\limits_{(x,\,y)\in\mathcal{S}} s_{x_k,y_k,z_k}^\ast(x,y)\left(r(x,y,t)\star\mathrm{sinc}_T(t)\Big|_{t=mT}\right)\mathrm{d}x \mathrm{d}y \notag \\
&=&\sum_{\ell=0}^{K-1}\sqrt{P_kP_\ell} u_\ell[m] \iint\limits_{(x,\,y)\in\mathcal{S}} s_{x_\ell,y_\ell,z_\ell}(x,y)s_{x_k,y_k,z_k}^\ast(x,y)\mathrm{d}x \mathrm{d}y+w_k[m] \notag  \\
&=&\sum_{\ell=0}^{K-1}\sqrt{P_kP_\ell} \phi_{k,\ell }u_\ell[m]+w_k[m], 
  \eea}
\hspace{-1.4mm}where $w_k[m]$ is the effective discrete noise after MF, and the coefficient
 \bea \label{phi} \phi_{k,\ell } =\iint\limits_{(x,\,y)\in\mathcal{S}}s_{x_\ell,y_\ell,z_\ell}(x,y) s^\ast_{x_k,y_k,z_k}(x,y)\mathrm{d}x \mathrm{d}y, \eea
where $\mathcal{S}$ is the surface-area spanned by the two-dimensional LIS. As the received signal after MF is identical for all samples and there is no temporal interference, we omit the index $m$ and assemble the notation in (\ref{rkm}) into a matrix formulation as
\bea \label{rk} \vec{r} =\vec{G}\vec{u} + \vec{w}, \eea
where the $(\ell, k)$th element of matrix $\vec{G}$ equals 
\bea \label{glk} g_{k,\ell}=\sqrt{P_\ell P_k}\phi_{k,\ell },\eea
which represents the received signal power when $k\!=\!\ell$ and the inter-user interference when $k\!\neq\!\ell$, respectively. Note that, with the MF process the noise variables are still zero-mean but colored with a covariance matrix 
\bea \mathbb{E}[\vec{w}\vec{w}^\mathrm{H}]=N_0\vec{G}.\eea
In the rest of this paper, we assume equal terminal transmit powers (per Hz) $P_k\!=\!P$ and study the capability of the terminals to communicate with the LIS.

\subsection{Independent Signal Dimensions with the LIS}

With the received ISI signal model (\ref{rk}), the channel capacity $\mathcal{C}$ averaged by the number of terminals, in nats per channel use, equals\cite{T99}
\bea \label{Cap} \mathcal{C}\!=\!\frac{1}{K}\log\left(\vec{I}+\frac{\vec{G}}{N_0}\right) .\eea
The capacity $\mathcal{C}$ is also identical to the capacity in nats/s/Hz by noting that $\mathcal{C}/(TW)\!=\!\mathcal{C}$. Hence, in the rest of the paper,  we always assume that $\mathcal{C}$ has the unit nats/s/Hz, and pay no attention to the properties of $\mathcal{C}$ on time or frequency domains. Bearing this in mind, we put an emphasis on the number of independent signal dimensions per deployed area-unit of the LIS that is possible to harvest, which is calculated based on the capacity normalized with the total deployed surface-area; a quantity we refer to as $\hat{\mathcal{C}}$. Therefore, the capacity $\hat{\mathcal{C}}$ has the unit nats/s/Hz/area-unit. This can be interpreted as the number of available signal space dimensions per area-unit, in perfect analogy with Shannon\rq{}s original ideas \cite{Shan49}. Reaching this space-normalized capacity $\hat{\mathcal{C}}$ in practice requires, of course, the $K$ terminals to be (i) sufficiently many, and (ii) favorably located in space.

Next we define $\hat{\mathcal{C}}$ in detail. For a one-dimensional terminal-deployment such as in Fig. \ref{fig2}, $K$ terminals are uniformly deployed along a line that is in parallel to the LIS and with a spacing $\Delta_x$ between two adjacent terminals. As $K\!\to\!\infty$, the length of the terminal-deployment $K \Delta_x\!\to\!\infty$. We then consider a rectangular shaped\footnote{The shape of the LIS becomes irrelevant when the surface-area is infinitely large.} LIS whose length grows at the same rate\footnote{In one-dimensional case, we consider normalizing the capacity $\mathcal{C}$ by the length of the LIS, i.e., $K\Delta_x$, for analyzing the number of independent signal dimensions $\rho$. Otherwise, if we normalized $\mathcal{C}$ by the surface-area, the number of independent signal dimensions $\rho$ approaches zero when the width of the LIS also goes to infinity.}. The space-normalized capacity $\hat{\mathcal{C}}$ [nats/s/Hz/m] is calculated as 
\bea \label{Cbar1}\hat{\mathcal{C}}=\lim_{K\to\infty}\frac{K\mathcal{C}}{K\Delta_x}=\lim_{K\to\infty}\frac{\mathcal{C}}{\Delta_x}. \eea

For two or three dimensional terminal-deployments where the spacings between two adjacent terminals are $\Delta x$ and $\Delta y$ for $x$ and $y$ dimensions\footnote{For three-dimensional case we omit the spacings between terminals in $z$-dimension as we can ideally project all the $K$ terminals to a $xy$-plane in front of the LIS for the purpose of analyzing the number of independent signal dimensions $\rho$, as what will become clear later in Sec. III-C.} such as in Fig. \ref{fig2}, respectively, we also consider a rectangular LIS whose surface-area grows at the same rate when $K\!\to\!\infty$. Denote $\Delta_s\!=\!\Delta_x\Delta_y$, the space-normalized capacity $\hat{\mathcal{C}}$ [nats/s/Hz/m$^2$] is calculated as
\bea \label{Cbar2}\hat{\mathcal{C}}=\lim_{K\to\infty}\frac{K\mathcal{C}}{K\Delta_x\Delta_y}=\lim_{K\to\infty}\frac{\mathcal{C}}{\Delta_s}.\eea

With $\hat{\mathcal{C}}$ defined in (\ref{Cbar1}) and (\ref{Cbar2}), respectively, the number of independent signal dimensions $\rho$ is calculated as the pre-log factor, i.e., the high signal-to-noise ratio (SNR) slope of $\hat{\mathcal{C}}$,
\bea \label{rho1} \rho = \lim_{\hat{P}/N_0\to \infty} \frac{\hat{\mathcal{C}}}{\log\left(\hat{P}/N_0\right)},\eea
where $\hat{P}$ is the transmit power (per Hz) per volume-unit of the terminal-deployments, i.e.,
\bea \label{con1} \hat{P}=\frac{P}{\Lambda},\eea
where $\Lambda\!=\!\Delta_x$ and $\Delta_s$ for one and two dimensional deployments, respectively. We point out that, $\hat{P}$ instead of $P$ shall be used in (\ref{rho1}) to calculate $\rho$. Otherwise, the normalized capacity $\hat{\mathcal{C}}$ becomes infinitely large when $\Lambda$ is small for a given $P$.

\subsection{Array Gain Considerations}
Let us first consider the received signal power (per Hz) at the LIS from an omni-directional antenna with power $P$ that is located at coordinates $x\!=\!y\!=\!0$ and $z\!=\!z_0$, that is, $z_0$ meters from the LIS and perpendicular to its center. The received power (per Hz) at the LIS, according to (\ref{phi}) and (\ref{glk}), equals
 \bea \label{Prx} g_{k,k} =P\iint\limits_{(x,\,y)\in\mathcal{S}}|s_{0,0,z_0}(x,y)|^2 \mathrm{d}x\mathrm{d}y = \zeta P,\eea
where
\bea \zeta= \frac{1}{4\pi}\iint\limits_{(x,\,y)\in\mathcal{S}}\frac{z_0}{\left(z_0^2+x^2+y^2\right)^{\frac{3}{2}}} \mathrm{d}x\mathrm{d}y.\eea
Assuming a rectangular LIS with $-A\!\leq \!x\!\leq\! A$ and $-B\!\leq\! y\!\leq\! B$, then $\zeta$ equals
\bea \label{nu} \zeta=\frac{1}{\pi}\tan ^{-1}\left(\frac{AB}{z_0\sqrt{A^2+B^2+z_0^2}} \right). \eea

Further, If one dimension of the LIS is much larger than the other dimension, e.g., $A\!\gg\!B$, the received power at the LIS reads
\bea \label{gkk} g_{k,k}= \frac{P}{\pi}\tan ^{-1}\!\left(\frac{B}{z_0} \right). \eea
Moreover, if both dimensions of the LIS are asymptotically large, i.e., $A\!=\!B\!=\!\infty$, then it holds that $g_{k,k}\!=\!P/2$, which makes intuitive sense, since half of the isotropically transmitted power from the terminal reaches the LIS, and the other half propagates away from. This number should now be compared to the free-space path-loss $\varepsilon_L$ that would result from a single receive antenna at distance $z_0$, which is typically many orders of magnitudes smaller than $P/2$. Thus we obtain, in addition to a possibly large value of independent signal dimensions, an impressive array gain.

\subsection{On the Approximation of an Integral} \label{Sec:approx}
As from (\ref{phi}), working with LIS results in solving an integral to calculate $\phi_{k,\ell}$. However, for the cases $\ell\!\neq\!k$, closed-form solutions seem out of reach and we seek for close approximations. We first state the following property that can be used to approximate $\phi_{\ell k}$.
\begin{property}
For sufficiently small $\lambda$, the integral
\bea \label{eq1} g(\Delta) \!=\! \int_{-\infty}^{\infty}\left(1+x^2\right)^{-\frac{3}{4}}\left(1+(x+\Delta)^2\right)^{-\frac{3}{4}} \exp\!\left(\!-\frac{2\pi\jmath}{\lambda}\left[\sqrt{1+x^2}\!-\!\sqrt{1+(x+\Delta)^2}\right]\right)\!\mathrm{d}x, \eea 
can be well approximated by a sinc function\footnote{The sinc-function without any subscript denotes a standard sinc-function with unit-energy and a double bandwidth 1.}
\be \label{eq2} g (\Delta) \approx 2\mathrm{sinc}\left(\frac{2\Delta}{\lambda}\right).\ee
\end{property}

\begin{figure}[t]
\begin{center}
\vspace*{-7mm}
\hspace*{-0mm}
\scalebox{.42}{\includegraphics{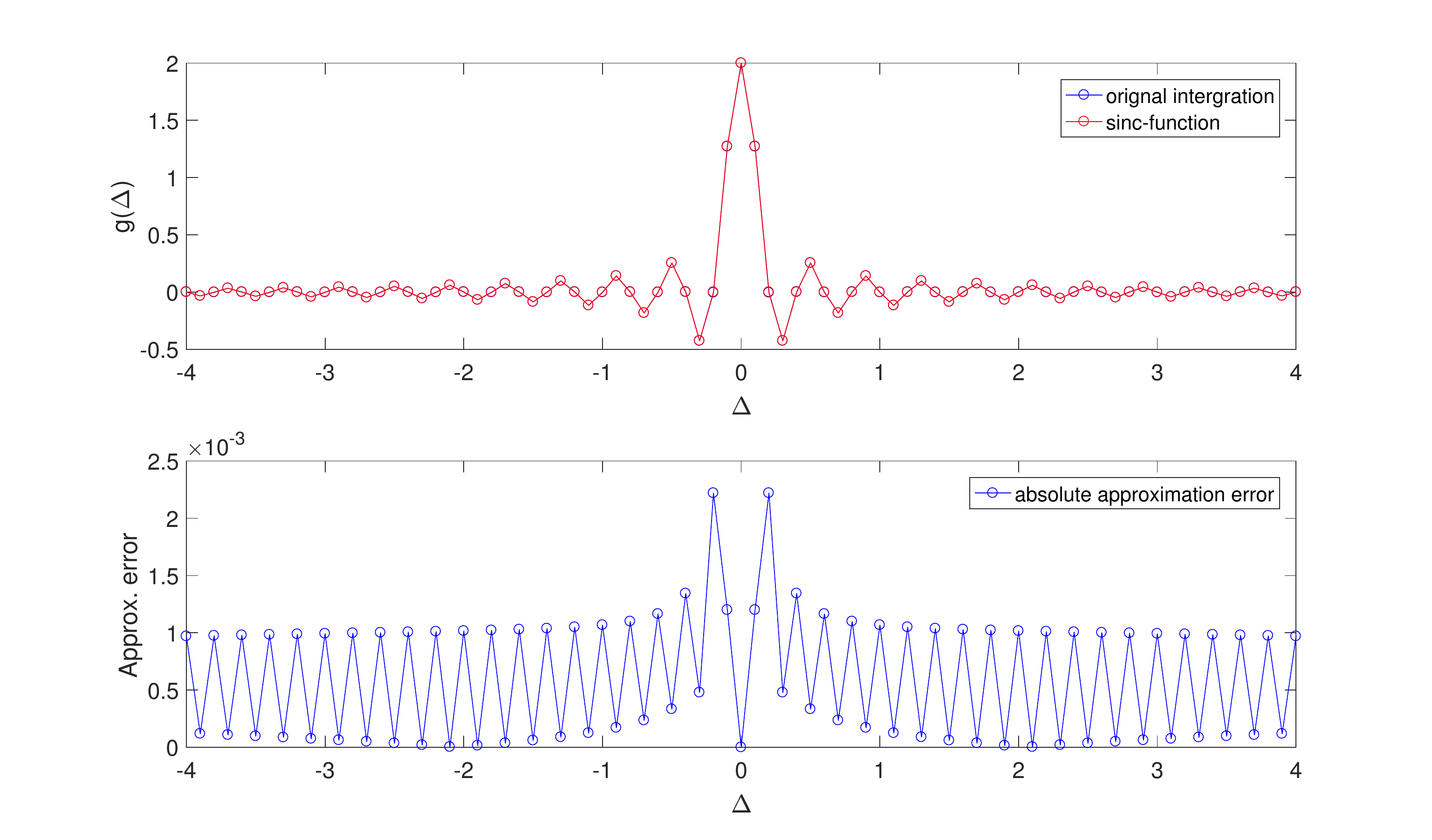}}
\vspace*{-10mm}
\caption{\label{fig3}The approximation of integration (\ref{eq1}) and the approximation errors for $g(\Delta)$ with $\lambda\!=\!0.4$ m. As can be seen, the errors are relatively small compared to the maximum value $g(0)$, i.e., the received power of the considered terminal.}
\vspace*{-12mm}
\end{center}
\end{figure}

Argumentations leading to Property 1 are in Appendix A. To answer what is meant by \lq\lq{}sufficiently small $\lambda$\rq\rq{}, we need also to take the distance $z_k$ from the terminal to the LIS into account. From the argumentation in Appendix A, it can be observed that $\lambda/z_k \!\lesssim\!1$. For wavelengths encountered in radio transmission, and reasonable distances from the surface, this condition is usually well satisfied. In Fig. \ref{fig3} we show an example of calculating $g(\Delta)$ with the exact integral (\ref{eq1}) and the approximation (\ref{eq2}) for $\lambda\!=\!0.4$ m. As can be seen, the two curves are almost aligned with each other and the approximation errors are relatively small. With (\ref{eq2}), we can then analyze the information-theoretical properties of the LIS in the next section.

\section{Space-normalized Capacities and Independent Signal dimensions} \label{Sec3}
In this section, we take an information-theoretical analysis on signal model (\ref{rk}) for one, two and three dimensional deployments of the $K$ terminals, and derive the number of independent signal dimensions that can be harvested with a LIS for a given transmit power per volume-unit.

\subsection{Capacity for One-Dimensional Case: Terminals on a Line}
We start with the one-dimensional terminal-deployment and consider an infinitely long LIS with a rectangular shape with $-\infty\!\leq\! x\!\leq\! \infty$ and $-B\!\leq\! y\!\leq\! B$, where terminals are uniformly located along the line with coordinates $y\!=\!0$ and $z\!=\!z_0$, with a spacing-distance $\Delta_x$ between two adjacent terminals\footnote{Although an infinitely long wall and equi-distant terminal locations are unreasonable in practice, these assumptions are made for analytical tractability. General capacity results will be obtained, from which the insights of general capacity behavior can be concluded. Numerical results on LIS with finite sizes and random terminal positions will be given in Sec. VI, which are shown to be well predicted by the theoretical analysis.} as shown in Fig. \ref{fig2}. For notational convenience, we first define the ratio between the half wavelength and the terminal-spacing in one-dimensional deployment as
 \bea \label{theta} \theta\!=\frac{\lambda}{2\Delta_x}. \eea
As will be seen later, $\theta$ plays a key role in the following analysis. 

From (\ref{rk}) the received signal at the LIS for the $k$th terminal can be expressed as
\bea \label{mod1} r_k =\sum_{\ell=0}^{K-1}g_{k,\ell}u_{\ell}+w_{k},\eea
where the noise variables $w_k$ are zero-mean Gaussian variables with variances $\mathbb{E}[w_{k}^{\ast}w_{\ell}]\!=\!N_0g_{k,\ell}$.
\begin{property}
Using Property 1, the effective channel impulse response $g_{k, \ell}$ is real and equals
\bea \label{gkl1} g_{k,\ell} =\frac{P}{\pi}\tan ^{-1}\left(\frac{B}{z_0} \right)\mathrm{sinc}\left(\frac{k-\ell}{\theta}\right).\eea
\end{property}
\begin{proof}
See Appendix B.
\end{proof}
Note that, when $k\!=\!\ell$, (\ref{gkl1}) is well aligned with (\ref{gkk}) and the approximation in Property 1 is in fact exact. 

We first consider applying an optimal receiver on signal model (\ref{mod1}), and the capacity [nats/s/Hz] of each terminal can be calculated as \cite{HR162, RP12}
 \bea \label{Copt} \mathcal{C}=\frac{1}{\theta}\int_{-\theta/2}^{\theta/2}\!\log\left(1+\frac{G(f)}{N_0}\right)\mathrm{d}f, \eea
where $G(f)$ is the frequency response of ISI channel $g_{k,\ell}$ in (\ref{mod1}). Since $g_{k,\ell}$ are discrete samples of the sinc-function at a sampling rate $\theta$, and by the Poisson summation formula \cite{BZ97}, $G(f)$ can be expressed as
\bea \label{Gf} G(f)= \zeta P \theta\sum_{k=-\infty}^{\infty}G_0(f-k\theta), \eea
and $G_0(f)$ is the standard rectangular function i.e., the Fourier transform of sinc$(x)$. 

Defining two useful auxiliary variables
\bea \label{a} \alpha=\frac{1}{\theta}-\beta \;\;\mathrm{and}\;\; \label{b} \beta=\left\lfloor \frac{1}{\theta}\right\rfloor\!, \eea
and with the definition in (\ref{con1}), the capacity (\ref{Copt}) for the one-dimensional terminal-deployment is stated in Property 3.
\begin{property} 
With an infinitely long LIS in the direction where the terminals are deployed along a line with equal spacing, the capacity, with an optimal receiver, for each terminal is
 \bea \label{Copt3} \mathcal{C}= \alpha\log\left(1+\frac{(\beta+1) \lambda\zeta\hat{P} }{2N_0}\right)+(1-\alpha)\log\left(1+\frac{\beta \lambda\zeta\hat{P} }{ 2N_0}\right)\!.\eea
 \end{property}
 \begin{proof}
 See Appendix C.
 \end{proof}  
Whenever $\alpha\!=\!0$, i.e., $1/\theta$ is an integer, from (\ref{Copt3}) the capacity equals 
 \bea  \label{optC} \mathcal{C}= \log\left(1+\frac{\zeta P}{N_0}\right) \eea
which is the resulting capacity of a terminal if no other terminals are present and with an SNR equal to $\zeta P/N_0$. This is so since under such cases, $g_{k,\ell}\!=\!0$ for $\ell\!\neq\!k$. We remark that the analysis and discrete-time model of the one-dimensional case is identical to that of a faster-than-Nyquist (FTN) signaling system using a sinc-pulse \cite{FTN1,FTN2}. \lq{}

With the capacity $\mathcal{C}$ given in Property 3, we can obtain the space-normalized capacity $\hat{\mathcal{C}}$ in (\ref{Cbar1}). By directly evaluating the limit when $\lambda\!\to\!0$, we have the below corollary.
\begin{corollary}
As $\lambda\!\to\!0$, for any $\theta$ (i.e., $\Delta_x$) the space-normalized capacity $\hat{\mathcal{C}}$ converges to $\zeta\hat{P} /N_0$ [nats/s/Hz/m].
\end{corollary}

Instead of using an optimal receiver, we also consider the MF capacity corresponding to model (\ref{mod1}), and we summarize our findings in Property 4.
 \begin{property} 
Under the same assumptions in Property 3, the capacity [nats/s/Hz] per-terminal with only the MF process applied in front is
 \bea \label{Cmf} \mathcal{C}=\log\left(1+\frac{\zeta P}{N_0+I}\right),  \eea
where the interference power $I$ equals
 \bea \label{I} I=\zeta P\bigg(\theta^2\left(\beta^2+2\alpha\beta+\alpha\right)-1\bigg). \eea
 \end{property} 
  \begin{proof} 
  See Appendix D.
\end{proof} 
   
From (\ref{I}), under the cases that $1/\theta$ is an integer, the interference power $I\!=\!0$ and the MF capacity (\ref{Cmf}) is the same as the capacity for the interference-free case. With the capacities $\mathcal{C}$ in (\ref{Copt3}), from (\ref{rho1}) it can be shown that with the optimal receiver,
\bea \rho =\left\{ \begin{array}{cc} 2/\lambda&\theta\geq1 \\ 2\theta/\lambda  &\mathrm{otherwise}. \end{array}  \right.  \notag\eea
Therefore, the maximal number of independent signal dimensions per m is $2/\lambda$ for one-dimensional terminal deployments. When $1/\theta$ is an integer (or when $\lambda$ is sufficiently small), the MF can also achieve the same same asymptotic slope of the normalized capacity curve $\hat{\mathcal{C}}$ from (\ref{Cmf}) .

\subsection{The Two-Dimensional Case: Terminals on a Plane} \label{Sec:plane}
We next move on to the case that, terminals are located on a two-dimensional plane at $z\!=\!z_0$ which is in parallel to the LIS plane as in Fig. \ref{fig2}. We are concerned with the number of independent signal dimensions per m$^2$ deployed surface-area, and we therefore let $A,B \!\to\! \infty$ to avoid edge effects. In this case, $\zeta\!=\!1/2$ for all $z_0$ and capacity does not depend on distance as a result of Property 2. 

The first step is to study the spatial PSD of received signal $r(x, y, t)$ in the absence of noise. Technically, we look at the received signal after sinc-based matched filtering only (i.e., the convolution in (\ref{rkm})), but not the spatial correlator (i.e. the integrals in (\ref{rkm})). The PSD is given by the two-dimensional Fourier transform \cite{TL93} of the autocorrelation 
\bea g(\Delta_x,\Delta_y) = \int_{-\infty}^\infty\int_{-\infty}^\infty s_{0,0,z_0}(x,y)s_{\Delta_x,\Delta_y,z_0}^\ast (x,y)\mathrm{d}x\mathrm{d}y. \eea
Note that, as the LIS is infinitely long in both dimensions, only the distance
\bea \tau=\sqrt{\Delta_x^2\!+\!\Delta_y^2}\eea
between two adjacent terminals matters for calculating $g(\Delta_x,\Delta_y) $. Under the approximation of Property 2, we have
\bea g(\tau) \approx \frac{1}{2} \mathrm{sinc}\left(\frac{2\tau}{\lambda}\right).\eea
As this function has radial symmetry, it follows that its Fourier transform is given by the Hankel transform \cite{H85} of degree zero, i.e.,
 {\setlength\arraycolsep{2pt}  \bea \label{Hankel} G(s) &=& 2\pi \mathcal{H}_0\left\{g(\tau)\right\} \nonumber \\
&=& \pi \int_{0}^\infty \!\!\tau\mathrm{sinc}\left(\frac{2\tau}{\lambda} \right){J}_0\!\left(2\pi s\tau\right)\!\mathrm{d}\tau\nonumber \\
&=& \left\{\begin{array}{ll} \frac{\lambda}{4\pi}\!\left(\sqrt{\frac{1}{\lambda^2}-s^2}\right)^{\!-1}, & 0\leq s < \frac{1}{\lambda} \\
		 0, & s>\frac{1}{\lambda} \end{array} \right. \eea}
\hspace{-1.8mm}where ${J}_{\,0}(x)$ is the zeroth-order Bessel function of the first kind \cite{AS72}. With the transmit power $\hat{P}$ per m$^2$ defined in (\ref{con1}), the space-normalized capacity $\hat{\mathcal{C}}$ [nats/s/Hz/m$^2$] equals
{\setlength\arraycolsep{2pt}  \bea \label{Cbar2D}  \hat{\mathcal{C}} &=&\int_{0}^{2\pi} \int_{0}^{1/\lambda} s\log\left(1+\frac{\hat{P}}{ N_0}G(s)\right) \mathrm{d}s\mathrm{d}\theta \nonumber \\
&=&\pi\left(\frac{\log(1+\lambda N)}{\lambda^2}+N^2\log\left(\frac{N\lambda}{1+N\lambda}\right)+\frac{N}{\lambda}\right)\!, \quad\eea}
\hspace{-1.4mm}where
$$N = \frac{\lambda \hat{P}}{4\pi N_0}.$$

By directly evaluating the limit of (\ref{Cbar2D}) as $\lambda\!\to\!0$, we obtain the below corollary.
\begin{corollary}
As $\lambda\!\to\!0$, the limit of space-normalized capacity $\hat{\mathcal{C}}$ equals $\hat{P}/2N_0$ [nats/s/Hz/m$^2$], which is the same as the one-dimensional case with $B\!=\!\infty$ in Corollary 1.
\end{corollary}
Then, the number of independent signal dimensions that can be harvested for the two-dimensional terminal-deployment can be computed by directly evaluating (\ref{rho1}) with $\hat{\mathcal{C}}$ in (\ref{Cbar2D}), which is stated in the below property.
\begin{property}
The number of independent signal dimensions for the two-dimensional terminal-deployment is
\bea \label{rho2} \rho = \frac{\pi}{\lambda^2}. \eea
Thus, for every $\lambda^2$ deployed surface-area of the LIS, we obtain $\pi$ independent signal dimensions.
\end{property}

\subsection{The Three-Dimensional Case: Terminals in a Sphere}
From the derivations for two-dimensional case in Sec. \ref{Sec:plane}, we have already furnished for a solution of the dimensionality for the three-dimensional terminal-deployment. Consider the Fourier transform $S_{x_0,y_0,z_0}(f_1,f_2)$ of a signal $s_{x_0,y_0,z_0}(x,y)$. 
From the convolutional property of Hankel transforms \cite{H85}, it follows that $G(s)$ in (\ref{Hankel}) is given by 
\bea \label{Gf1f2} G\left(\!\sqrt{f_1^2+f_2^2}\right)\! = \big|S_{x_0,y_0,z_0}(f_1,f_2)\big|^2. \eea
Since $G(s)$ in (\ref{Hankel}) does not depend on $z_0$, (\ref{Gf1f2}) implies that, the domain of $S_{x_0,y_0,z_0}(f_1,f_2)$ is independent of the distance $z_0$ from the wall. Since the number of independent signal dimensions that can be accommodated is proportional to the area of the domain of $S_{x_0,y_0,z_0}(f_1,f_2)$, it follows that the same number of dimensions is obtained in the three-dimensional case as in the two-dimensional case.

An alternative way to realize this result is to consider a hyper plane $\mathcal{P}=\{x,y,z:z\!=\!z_0\}$ for some small $z_0$. All signals transmitted from terminals at $z_{k}\!>\!z_0$ has to pass the plane $\mathcal{P}$. From the Huygens-Fresnel principle \cite{BW80} it, however, follows that the signal that reaches the LIS can be expressed as point sources at the plane $\mathcal{P}$ that radiate the signals arrived $\mathcal{P}$ from the terminals. However, the number of signal space dimensions at the plane $\mathcal{P}$ is $\pi/\lambda^2$ per m$^2$, which means that the number of dimensions in the three-dimensional volume is unaltered compared to the two-dimensional case.

\section{Implementing the LIS based on Sampling Theory}
We have seen in Sec. \ref{Sec3} that, the received signal at the LIS has a two-dimensional Fourier transform that is band-limited to a disc of radius $1/\lambda$. A direct consequence is that, there is no loss if the received signal $\tilde{s}_{x_0,\,y_0,\,z_0}(x,y)$ is sampled sufficiently dense so that no aliasing occurs. Thus, a LIS can be implemented as a grid of discrete antenna-elements. In this section we take a look at optimal sampling of the LIS. As we deal with LISs with unbounded physical dimensions, we make use of lattice theory \cite{PM62, KA05, HM91}.

We can foresee at least two objectives for designing the sampling. (i) One view of optimal sampling is that the antenna units are costly, while area is available in excess. With this view, we should constrain ourselves to obtain one signal space dimension for every spent antenna. Once this constraint is met, we should then find the sampling lattice that minimizes the area. (ii) An alternative path would be, as established in Sec. \ref{Sec3}, a LIS can at most offer $\pi/\lambda^2$ dimensions per m$^2$ deployed surface-area, one can constrain the sampled LIS to offer the same number of dimensions per m$^2$, and then to ask for the least number of antenna elements per m$^2$ that meets the constraint. With this view, the physical resource to be minimized is area, while antenna units are considered cheap. Problems (i) and (ii) are similar, and in this paper we treat (i) in depth. We point out that for (ii), the resulting lattice problem to be faced is to find the densest lattice whose fundamental cell circumscribes a circle of given radius \cite{KH05}.

Suppose that sampling of the LIS is made on the basis of the sampling matrix $\vec{S}$. With that, the asymptotic number of placed antennas per m$^2$ is $A_{\mathrm{d}}=1/|V(\vec{S})|$, where $V(\vec{S})$ is the fundamental cell of a lattice generated from $\vec{S}$, and $|V(\vec{S})|$ is its fundamental volume. Let $\tilde{s}_{m,n}$ denote the samples of $\tilde{s}_{x_0,\,y_0,\,z_0}(x,y)$ generated from sampling at 
$$\left[\begin{array}{c} x \\ y\end{array} \right] = \vec{S}\left[\begin{array}{c} m \\ n\end{array} \right].$$ The samples $\{\tilde{s}_{m,n}\}$ have a Fourier transform $\tilde{{S}}(f_1,f_2)$ that is defined on the fundamental volume $V(\vec{S}^{-T})$ corresponding to the reciprocal lattice and is, possibly, an aliased version of $\tilde{S}(f_1,f_2)$. The capacity per antenna for the sampled LIS is then given by\cite[Theorem II.2]{CS03}
\be \label{lattice} {C}_{\mathrm{ant}} = \frac{1}{|V(\vec{S}^{-T})|}\int_{V(\vec{S}^{-T})}\log\left(1+\frac{\hat{P}}{|V({\vec{S}})| N_0}|\tilde{S}(f_1,f_2)|^2 \right)\mathrm{d}f_1\mathrm{d}f_2.\ee
However, recalling the bandlimited structure of $S_{x_0,\,y_0,\,z_0}(f_1,f_2)$, the support of $\tilde{S}(f_1,f_2)$ is limited to $D(\lambda^{-1}) \cap V(\vec{S}^{-T}),$ where $D(\lambda^{-1})$ denotes a disc with radius $1/\lambda$. Thus, we have that 
\bea \label{lattice} {C}_{\mathrm{ant}} = \frac{1}{|V(\vec{S}^{-T})|}\int_{D(\lambda^{-1}) \cap V(\vec{S}^{-T})}\log\left(1+\frac{\hat{P}}{|V({\vec{S}})| N_0}|\tilde{S}(f_1,f_2)|^2 \right)\mathrm{d}f_1\mathrm{d}f_2.\eea
The number of dimensions that we can harvest per spent antenna is defined, similar to (\ref{rho1}), as
\bea \rho_{\mathrm{ant}}=\lim_{\hat{P}/N_0 \to\infty} \frac{{C}_{\mathrm{ant}}}{\log(\hat{P}/N_0)}.\eea
It can be readily verified that this limit is given by
\bea \rho_{\mathrm{ant}} = \frac{|V(\vec{S}^{-T})|}{|D(\lambda^{-1}) \cap V(\vec{S}^{-T})|},\eea
and attains a maximum value $\rho_{\mathrm{ant}}\!=\!1$ whenever $V(\vec{S}^{-T}) \subset D(\lambda^{-1})$.

Let us now return to path (i) above. To satisfy the constraint of a maximum number of harvested dimensions per spent antenna, we know that we must sample with a lattice generator $\vec{S}$ satisfying  $V(\vec{S}^{-T}) \subset D(\lambda^{-1})$. To then minimize area, we should choose the lattice generator $\vec{S}$ such that its fundamental volume is minimized, i.e., the problem to be addressed is
{\setlength\arraycolsep{2pt} \bea \label{latticeproblem} && \min_{\vec{S}} |V(\vec{S})| \notag\\
&& \text{such that} \;\, V(\vec{S}^{-T}) \subset D(\lambda^{-1}).\eea}
\hspace{-1.4mm}We summarize the solution to this problem in the following property.
\begin{property}
The lattice generator that solves (\ref{latticeproblem}) is the scaled hexagonal lattice generator
$$\vec{S} = \left[ \begin{array}{cc} \frac{2\lambda}{{3}} & \frac{\lambda}{{3}} \\ 0 & \frac{\lambda}{\sqrt{3}} \end{array}\right].$$
\end{property}
\begin{proof}
See Appendix E.
\end{proof}

For this generator we have $\rho_{\mathrm{ant}}\!=\!1$ and an antenna density per m$^2$ that equals 
\bea \label{Ad1} A_{\mathrm{d}}=\frac{1}{\det(\vec{S})} = \frac{3\sqrt{3}}{2\lambda^2}.\eea
Let us now compare the hexagonal lattice with a LIS sampling according to a rectangular lattice generator. The most natural one to choose would perhaps be the generator
$\vec{S}\!=\!\!\left[ \begin{array}{cc} \frac{\lambda}{{2}} & 0\\ 0 & \frac{\lambda}{2} \end{array}\right]\!,$
which fails to meet the constraint $V(\vec{S}^{-T}) \subset D(\lambda^{-1})$. In fact, the generator for the densest rectangular lattice that meets the constraint, and thus results in $\rho_{\mathrm{ant}}\!=\!1$, is
$\vec{S}\!=\!\! \left[ \begin{array}{cc} \frac{\lambda}{\sqrt{2}} & 0\\ 0 & \frac{\lambda}{\sqrt{2}} \end{array}\right]\!.$ However, for this lattice generator we have $A_{\mathrm{d}}\!=\! 2/\lambda^2$. Combining this with (\ref{Ad1}) for the hexagonal case above, we can see that a hexagonal sampling of the LIS requires only a fraction $\frac{4}{3\sqrt{3}}\! \approx \!0.77$ of the surface-area required with a rectangular sampling, that is, 23\% of the surface-area can be saved. Note that the efficiency gain of the hexagonal sampling exceeds the normal packing gain $\frac{\pi}{2\sqrt{3}}\!\approx\!0.91$ of the hexagonal lattice, this being a result of our constraint that for each spent antenna, we should be able to harvest one signal space dimension. 

\section{Channel Shortening Demodulator Design for Data-Transmission with LIS}
In Sec. III we have derived the space-normalized capacity and independent signal dimensions with optimal receiver applied. Due to the prohibitive complexity of optimal detection algorithms such as BCJR \cite{BCJR}, reduced-complexity detectors are usually applied in practical systems. At the other end of the detection algorithm complexity-spectrum, one has the LMMSE equalizer. To reach the capacity $\mathcal{C}$, the BCJR is needed, while the LMMSE renders a loss. It is of interest to study this loss, but also to study detection algorithms that can operate with a computational complexity that falls in between the two extremes. One such class is channel shortening CS demodulators \cite{RP12, HR161, HR17j1}. While the literature on low-complexity detection algorithm is vast, the CS demodulator is chosen here since it is conceptually simple and can control its complexity via single variable $\nu$ that represents the memory depth of the search tree. 

Therefore, in this section we consider a low-complexity CS demodulator \cite{RP12, HR161, HR17j1} design for the considered LIS system for data-transmissions in the UL\footnote{Note that, CS based precoder designs can also be used to approach the DPC \cite{C83} capacity in DL as shown in \cite{HR17j2, HR173}.}. Although there has been no traditional channel matrix discussed in the paper so far, we introduce the main idea of CS as a receiver front-end that transforms an arbitrary channel matrix into the following form,
{\setlength\arraycolsep{1pt} \bea \label{Hmat} \vec{H}\!=\!\left[\!\begin{array}{ccccccc}
h_{1,1}&~&~&~&~&~&~\\ h_{2,1}& h_{2,2}&~&~&~&~&~\\  
\vdots&h_{3,2}&\ddots&~&~&~&~ \\ 
h_{\nu+1,1}&\vdots&\ddots&\ddots&~&~&~\\~&h_{\nu+2,2}&\ddots&\ddots&\ddots&~&~\\
~&~&\ddots&\ddots&\ddots&\ddots&~\\ ~&~&~&h_{K, K-\nu}&\cdots&h_{K, K-1}&h_{K, K}
\end{array} \!\right]\!\!, \;\;\eea}
\hspace{-1.4mm}where variable $\nu$ denotes the interfering depth among different terminals after CS process. 

CS demodulator has several advantages. Firstly, it provides a flexibility to switch between the optimal BCJR \cite{BCJR} and the linear MMSE demodulators \cite{MS02}. Secondly, with a small $\nu$ it can perform close to the optimal demodulator, yet with a much less complexity. Lastly, it has closed-form solutions for the CS parameters as what will explained in detail in the next. 

The CS demodulator comprises two steps: shortening the ISI channel matrix $\vec{G}$ into an effective channel $\vec{H}$ that has a shape\footnote{Conceptually, more precise descriptions can be found in \cite{HR17j1}.} of (\ref{Hmat}), followed by a BCJR demodulator with a number of states $|\mathcal{X}|^{\nu}$, where $|\mathcal{X}|$ is the size of the transmitted symbol-alphabet of each terminal, which without loss of generality here is assumed to be identical for all $K$ terminals. Compared to the optimal BCJR demodulator which has a number of states $|\mathcal{X}|^{K\!-\!1}$, the number of states is significantly reduced with a small $\nu$. The performance of CS demodulator will be given in Sec. VI, and in the rest of this section we briefly lay down the fundamentals of a CS demodulator and the optimization of its parameters. 

With the input-output relation of the transceiver given in (\ref{rk}), the CS demodulator operates on the Ungerboeck \cite{CB05} model $\tilde{p}(\vec{r}|\vec{u})$ that is defined as
\bea \label{modelT} \tilde{p}(\vec{r}|\vec{u}) = \exp\left(-2\mathcal{R}\left\{\vec{u}\rmh\vec{W}\rmh\vec{r}\right\}+\vec{u}\rmh\vec{ \Phi}\vec{u}\right),\eea
instead of the true conditional probability,
\bea p(\vec{r}|\vec{u}) = \frac{1}{(\pi N_0)^{K}\det(\vec{G})}\exp\left(\!-\frac{\left(\vec{r}-\vec{G}\vec{u}\right)\rmh\vec{G}^{-1}\left(\vec{r}-\vec{G}\vec{u}\right)}{N_0}\!\right)\!. \eea
The parameters $\vec{W}$ and $\vec{\Phi}$ in (\ref{modelT}) are optimized via maximizing the achievable information rate (AIR) \cite{HR17j1} that equals
\bea  \label{metric} I_{\mathrm{AIR}}\left(\vec{r};\vec{u}\right) =\mathbb{E}_{\vec{u},\vec{r}}\left[\log\tilde{u}(\vec{r}|\vec{u})\right]-\mathbb{E}_{\vec{r}}\left[\log \tilde{p}(\vec{r})\right].\eea
With CS demodulator, the matrix $\vec{\Phi}$ is constrained to be a band-shaped Hermitian matrix such that, only the middle $2\nu\!+\!1$ diagonals can take non-zero values with a Cholesky-decomposition
\bea\label{matPh}\vec{I}+\vec{\Phi}=\vec{H}\vec{H}\rmh,\eea
where $\vec{H}$ is a $K\!\times\! K$ lower-triangular effective channel matrix that is obtained through maximizing (\ref{metric}) under the constraint that is has the band-limited shape of (\ref{Hmat}).

Denote $\vec{B}$ as the minimum-mean-square-error (MMSE) matrix
\bea  \label{mse} \vec{B}=\left(\frac{\vec{G}\rmh}{N_0}+\vec{I}\right)^{\!-1},\eea
and introduce the notations
 {\setlength\arraycolsep{2pt}\bea \label{bk} \vec{b}_k^{\nu}&=&\left[b_{k, k+1}\;, b_{k, k+2}\;,\cdots\;,b_{k,k\boxplus\nu}\right]\!, \notag \\
\label{hk} \vec{h}_{k}^{\nu}&=&\left[h_{k+1,k}\;,h_{k+2,k}\;,\cdots\;,h_{k\boxplus\nu,k}\right]\rmt,\eea}
\hspace{-1.4mm}with the operation
 \bea k\boxplus\nu=\min(k+\nu, K-1).   \eea
Further, define the principle sub-matrix $\vec{B}_k^{\nu}$ obtained from $\vec{B}$ as
 {\setlength\arraycolsep{5pt} \bea \label{Bn} \vec{B}_{k}^{\nu}\!=\!\left[\begin{array}{cccc}  b_{k+1,k+1}& b_{k+1,k+2}&\cdots&b_{k+1,k\boxplus\nu}\\b_{k+2,k+1}& b_{k+2,k+2}&\cdots&b_{k+2,k\boxplus\nu}\\ \vdots&\vdots&\vdots&\vdots \\b_{k\boxplus\nu,k+1}&b_{k\boxplus\nu,k+2}&\cdots&b_{k\boxplus\nu,k\boxplus\nu}\end{array} \right]\!\!,  \eea}
\hspace{-1.4mm}Then, following the same approach as in \cite{RP12, HR161, HR17j1}, the optimal $K\!\times\! K$ prefiltering matrix $\vec{W}$ is
\bea \label{wopt} \vec{W}=\left(\vec{G}\rmh+ N_0\vec{I}\right)^{-1}\left(\vec{I}+\vec{\Phi}\right),\eea
and a closed-form for the optimal $I_{\mathrm{AIR}} $ in (\ref{metric}) can be reached as
\bea \label{AIR2} I_{\mathrm{AIR}}\left(\vec{r};\vec{u}\right) =2\sum_{k=0}^{K-1}\log h_{k,k},\eea
where the optimal diagonal elements of $\vec{H}$ are calculated as
 \bea \label{ukk}  h_{k,k}=\left\{\begin{array}{ll}\sqrt{\left(b_{k,k}- \vec{b}_k^{\nu}\left(\vec{B}_{k}^{\nu}\right)^{-1}\left(\vec{b}_k^{\nu}\right)\rmh\right)^{-1}},&\; 0\leq k<K-1, \\
\frac{1}{\sqrt{b_{K-1,K-1}}},& \;k=K-1,  \end{array}\right.  \eea
and non-diagonal elements of each column are equal to
\bea \vec{h}_{k}^{\nu}=-h_{k,k}\left(\vec{B}_{k}^{\nu}\right)^{-1}\left(\vec{b}_k^{\nu}\right)\rmh.  \eea

Under the two extreme cases $\nu\!=\!0$ and $\nu\!=\!K\!-\!1$, as shown in \cite{HR17j1,HR161} the CS demodulator is identical to the LMMSE and BCJR demodulators, respectively. The Ungerboeck model based branch metric in the BCJR algorithm of the CS demodulator is computed as
\bea \gamma(u_k, u_{k-1},\cdots, u_{k-\nu})\!\propto\!-2\mathcal{R}\left\{\!u_k^{\ast}\!\left(\tilde{r}_k\!-\!\sum_{\ell=1}^{\nu}\phi_\ell u_{k-\ell}\!\right)\!\right\}\!+\!\phi_0 |u_k|^2,\eea
for transmitted symbols $(u_k, u_{k-1},\cdots, u_{k-\nu})$, where $\vec{\phi}\!=\![\phi_0,\;\phi_1,\;\cdots,\;\phi_{\nu}]$ is the first $\nu\!+\!1$ non-zero elements in the first column of $\vec{\Phi}$, and $\tilde{r}_k$ is the received data of the $k$th terminal after prefiltering with $\vec{W}$, that is, the operation $\tilde{\vec{r}}\!=\!\vec{W}\rmh\vec{r}$.

\section{Numerical Results}
In this section, we present simulation results to illustrate the capacities, independent signal dimensions, and AIR of CS demodulators for data-transmission with the LIS as discussed in previous sections. In what follows, the mentioned values of noise PSD and transmit powers are linear.

\begin{figure}[t]
\begin{center}
\vspace*{-4mm}
\hspace*{-2mm}
\scalebox{.42}{\includegraphics{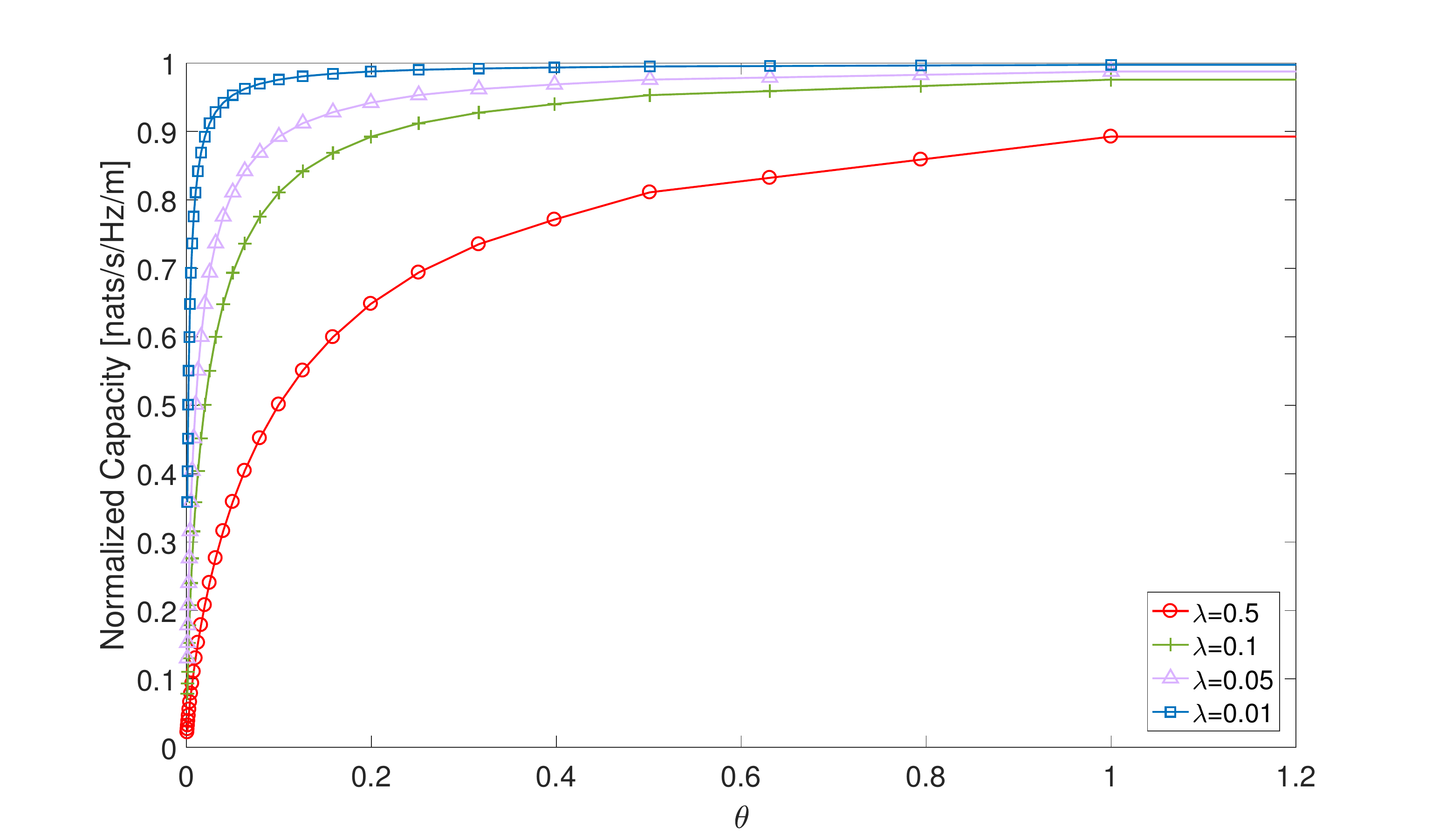}}
\vspace*{-7mm}
\caption{\label{fig4}The space-normalized capacity $\hat{\mathcal{C}}$ in relation to $\theta$ for optimal receiver with $N_0\!=1$, $\!\eta\!=\!0.1$, and $\hat{P}\!=\!10$.}
\vspace*{-7mm}
\end{center}
\end{figure}

\begin{figure}
\begin{center}
\vspace*{-4mm}
\hspace*{-2mm}
\scalebox{.42}{\includegraphics{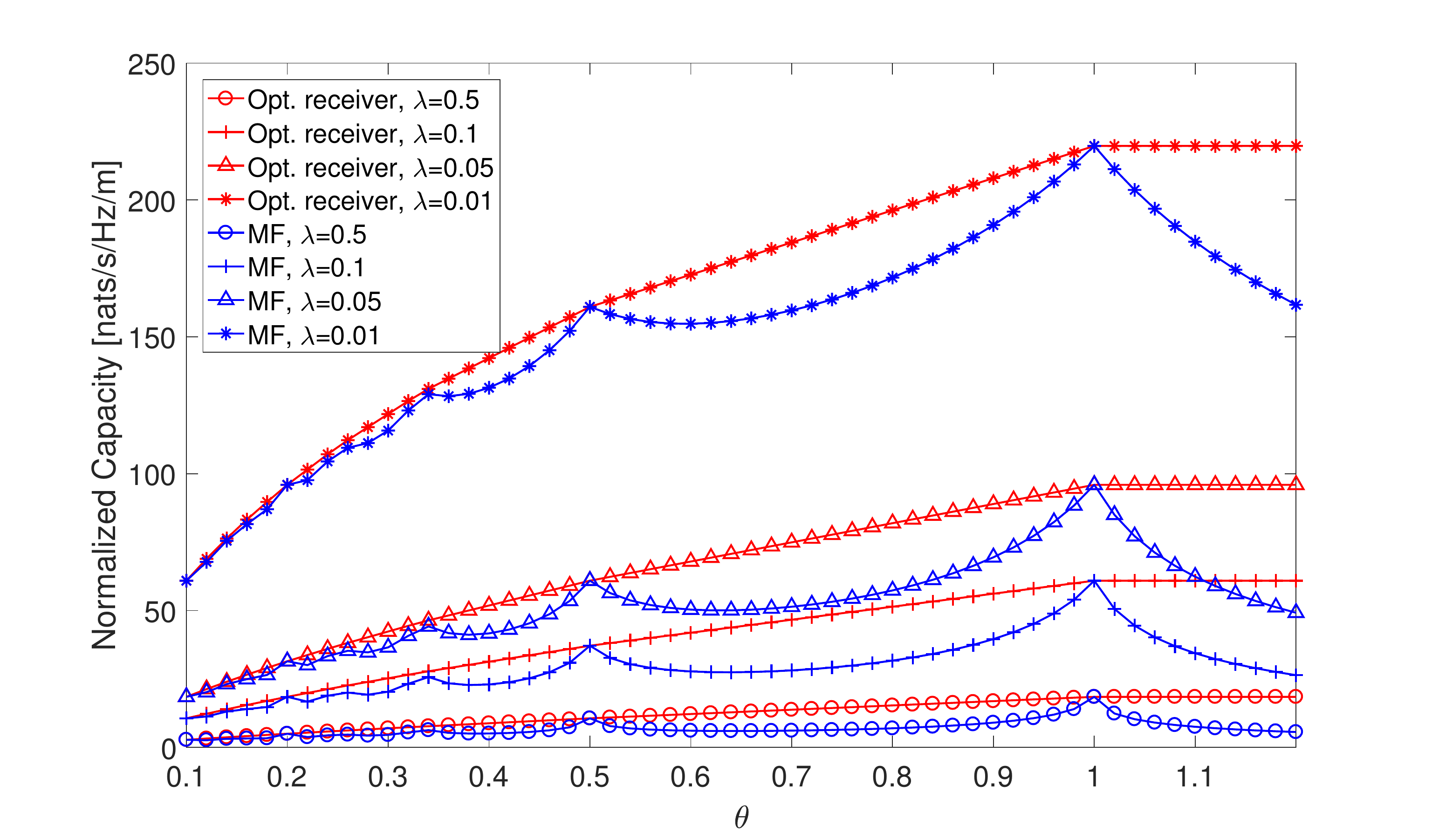}}
\vspace*{-7mm}
\caption{\label{fig5} The space-normalized capacity $\hat{\mathcal{C}}$ for the optimal and the MF receivers with $N_0\!=0.05$, $\!\eta\!=\!0.5$, and $\hat{P}\!=\!40$.}
\vspace*{-12mm}
\end{center}
\end{figure}

\begin{figure}[t]
\begin{center}
\vspace*{-4mm}
\hspace*{-2mm}
\scalebox{.42}{\includegraphics{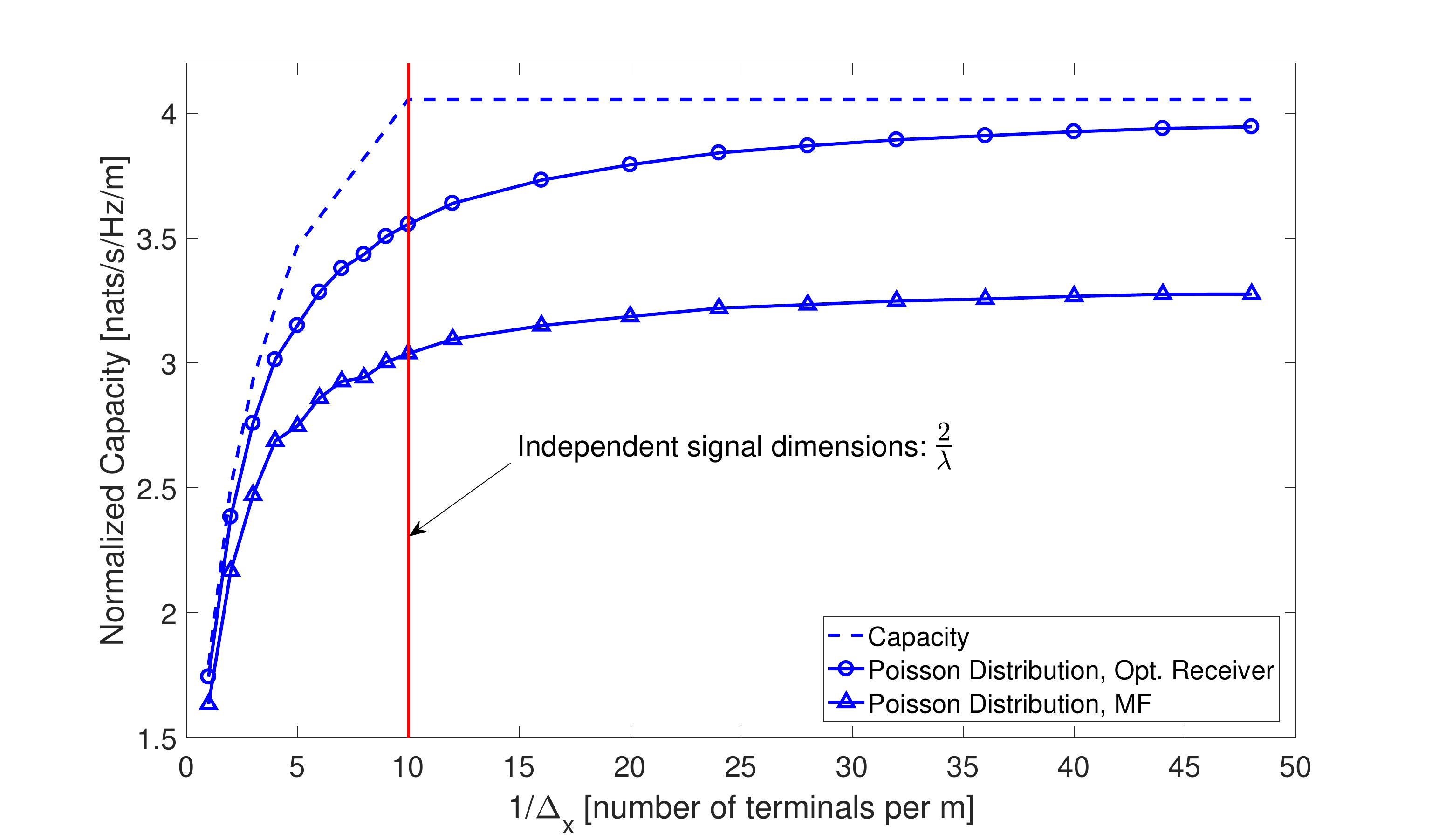}}
\vspace*{-7mm}
\caption{\label{fig7}The space-normalized capacity $\hat{\mathcal{C}}$ with randomly located terminals along a line with length 10 m. We assume that $A\!=\!B\!=\!\infty$, $N_0\!=\!1$, $\hat{P}\!=\!10$ and $\lambda\!=\!0.2$ m.}
\vspace*{-7mm}
\end{center}
\end{figure}

\begin{figure}
\begin{center}
\vspace*{-4mm}
\hspace*{-2mm}
\scalebox{.42}{\includegraphics{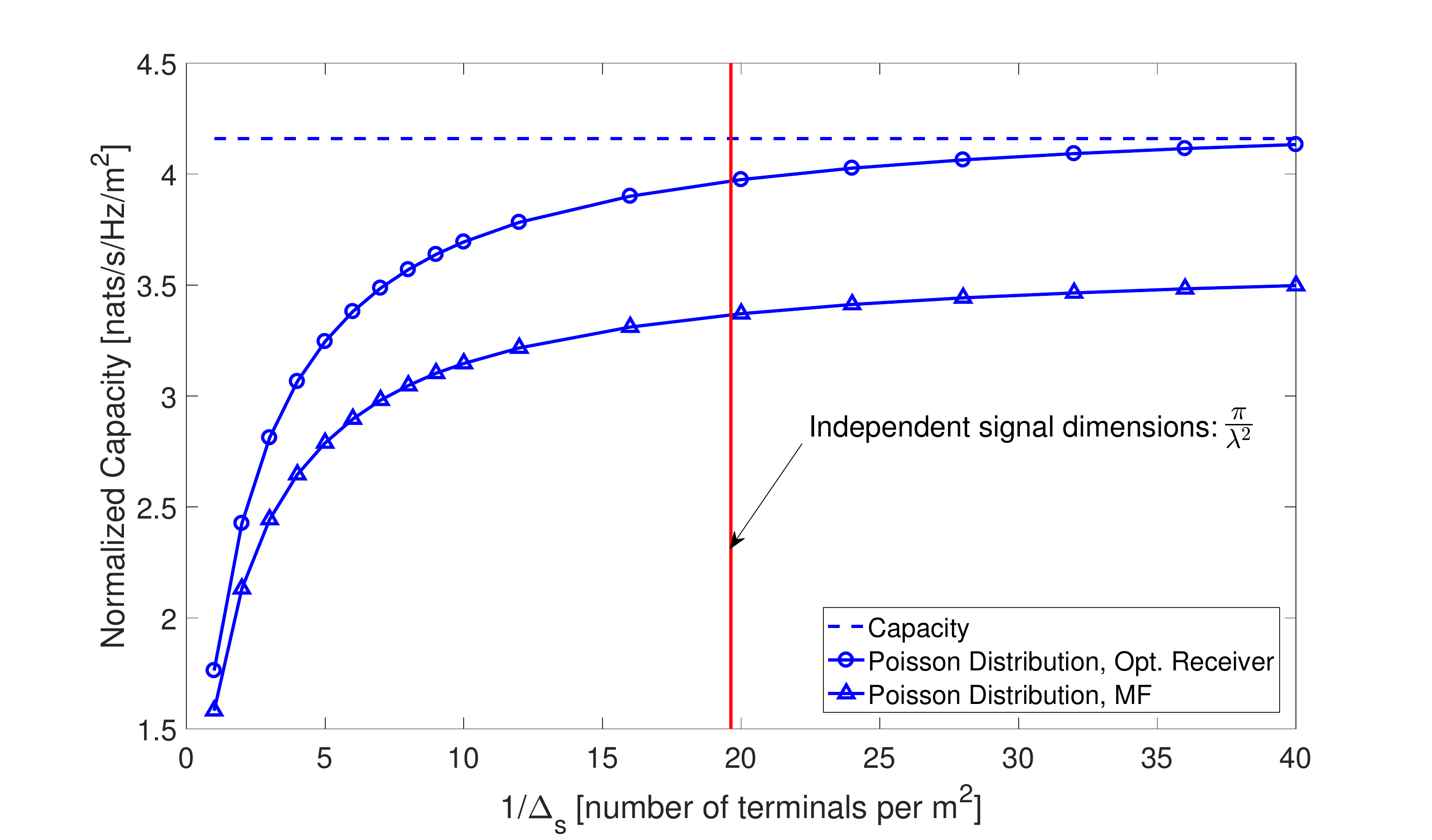}}
\vspace*{-7mm}
\caption{\label{fig8}The space-normalized capacity $\hat{\mathcal{C}}$ with randomly located terminals on a plane with both length and width equal to 20 m. We assume that $A\!=\!B\!=\!\infty$, $N_0=1$, $\hat{P}\!=\!10$ and $\lambda\!=\!0.4$ m.}
\vspace*{-12mm}
\end{center}
\end{figure}

\subsection{Capacities with LIS for One-Dimensional Terminal-deployments}

In Fig. \ref{fig4}, we evaluate the space-normalized capacity $\hat{\mathcal{C}}$ for one-dimensional terminal-deployment. We plot $\hat{\mathcal{C}}$ [nats/s/Hz/m] for $N_0\!=1$, $\!\zeta\!=\!0.1$, and $\hat{P}\!=\!10$, but for different values of $\Delta_x$ and $\lambda$ with the optimal receiver. As can be seen, as $\lambda\!\to\! 0$, $\hat{\mathcal{C}}$ converges to the limit $1$, which is aligned with Corollary 1.

In Fig. \ref{fig5}, we evaluate the differences in space-normalized capacities $\hat{\mathcal{C}}$ between the optimal and the MF receivers for $N_0\!=0.05$, $\!\zeta\!=\!0.5$, $\tilde{P}\!=\!40$, and for different values of $\Delta_x$ and $\lambda$. As can be seen, whenever $1/\theta$ is an integer, terminals do not interfere with each other and the normalized capacities $\hat{\mathcal{C}}$ of the optimal and the MF receivers are identical. Otherwise, the MF receiver is inferior to the optimal receiver as expected.

In Fig. \ref{fig7}, we evaluate the space-normalized capacity $\hat{\mathcal{C}}$ for uniformly random allocated terminals along a 10 m long line with different values of $\Delta_x$, with $1/\Delta_x$ representing the density of random allocations, i.e., in $L$ meters, we have $L/\Delta_x$ users randomly located. As can be seen, as $\Delta_x$ decreases to 0, the space-normalized capacity $\hat{\mathcal{C}}$ reaches the capacity limit with the optimal receiver and starts to saturate at $\Delta_x\!=\!\lambda/2\!=\!0.1$ m. With the MF receiver, the capacity also converges but is suboptimal.

\subsection{Capacities with LIS for Two and Three Dimensional Terminal-deployments}
Next, we evaluate the capacities for two and three dimensional terminal-deployments. In Fig. \ref{fig8}, we evaluate the space-normalized capacity $\hat{\mathcal{C}}$ for randomly located terminals in a two-dimensional plane with length and width both equal to 20 m. The locations of terminals are also drawn from a uniform distribution for a given terminal-density $1/\Delta_s$. As can be seen, when $\Delta_s$ decreases to 0, $\hat{\mathcal{C}}$ reaches a limit and starts to saturate at $\Delta_s\!=\!\lambda^2/\pi$ with the optimal receiver. With the MF receiver, the capacity also converges but is inferior to the optimal receiver similar to the conclusions drawn from the one-dimensional case.

\begin{figure}[t]
\begin{center}
\vspace*{-4mm}
\hspace*{-2mm}
\scalebox{.42}{\includegraphics{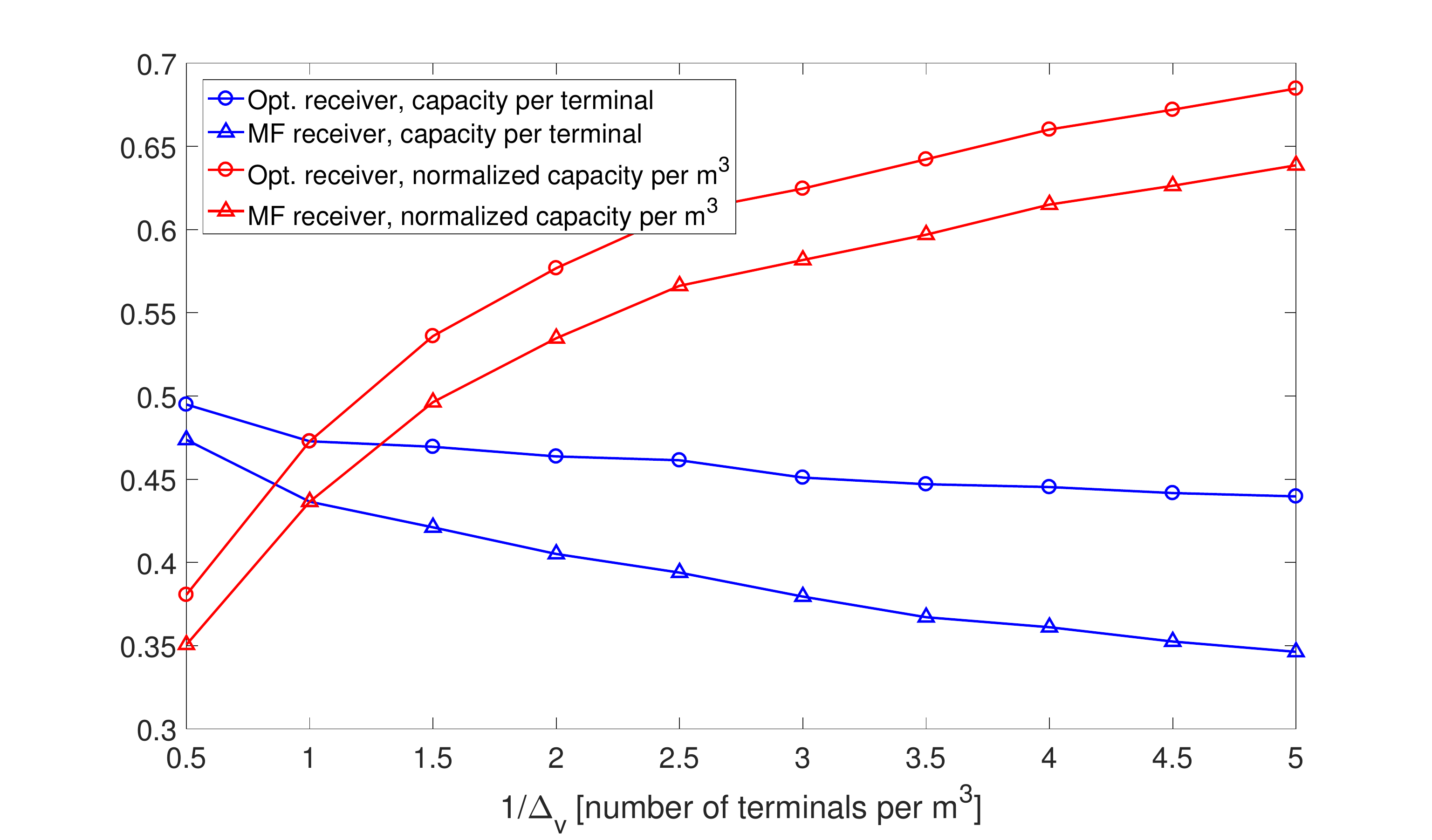}}
\vspace*{-7mm}
\caption{\label{fig9}The space-normalized capacity $\hat{\mathcal{C}}$ of randomly located terminals in a room with length, width and height are all equal to 4 m. We assume that $A\!=\!2$ m, $B\!=\!1$ m, $N_0\!=\!1$, $\hat{P}\!=\!10$ or $P\!=\!10$ and $\lambda\!=\!0.5$ m.}
\vspace*{-12mm}
\end{center}
\end{figure}

In Fig. \ref{fig9}, we evaluate the three-dimensional case, where we consider a room with length, width and height all equal to 4 m. For simplicity, we do not account for any reflections. On the front wall of the room, we assume a rectangular LIS, with length 2 m and width 1 m, deployed in the middle. For instance, we can use a white-board in a room as the LIS. Since we have a LIS with finite size, we use numerical computations to calculate the elements $\phi_{k,\ell}$ instead of using the sinc-function approximation. We evaluate the space-normalized capacity $\hat{\mathcal{C}}$ for randomly located terminals drawn from a uniform distribution in the room with different terminal-density $1/\Delta_v$, and consider two different cases. The first case is that, we fix the transmit power of each terminal to be $P\!=\!10$ and then measure the capacity $\mathcal{C}$ per terminal. The other case is that, we fix the transmit power per m$^3$ to $\hat{P}\!=\!P/\Delta_v\!=\!10$ (similar as the definition in (\ref{con1})) and estimate the space-normalized capacity $\hat{\mathcal{C}}$ per m$^3$. As can be seen, when $\Delta_v$ decreases to 0, $\hat{\mathcal{C}}$ increases both for the optimal and MF receivers, like in the one and two dimensional cases. The capacity $\mathcal{C}$, however, is fairly flat when the number of terminals increases from 32 to 320, while the latter one results in more interferences among terminals. This clearly shows the potential of the LIS for interference suppression in data-transmission.

\begin{figure}[t]
\begin{center}
\vspace*{-4mm}
\hspace*{-2mm}
\scalebox{.42}{\includegraphics{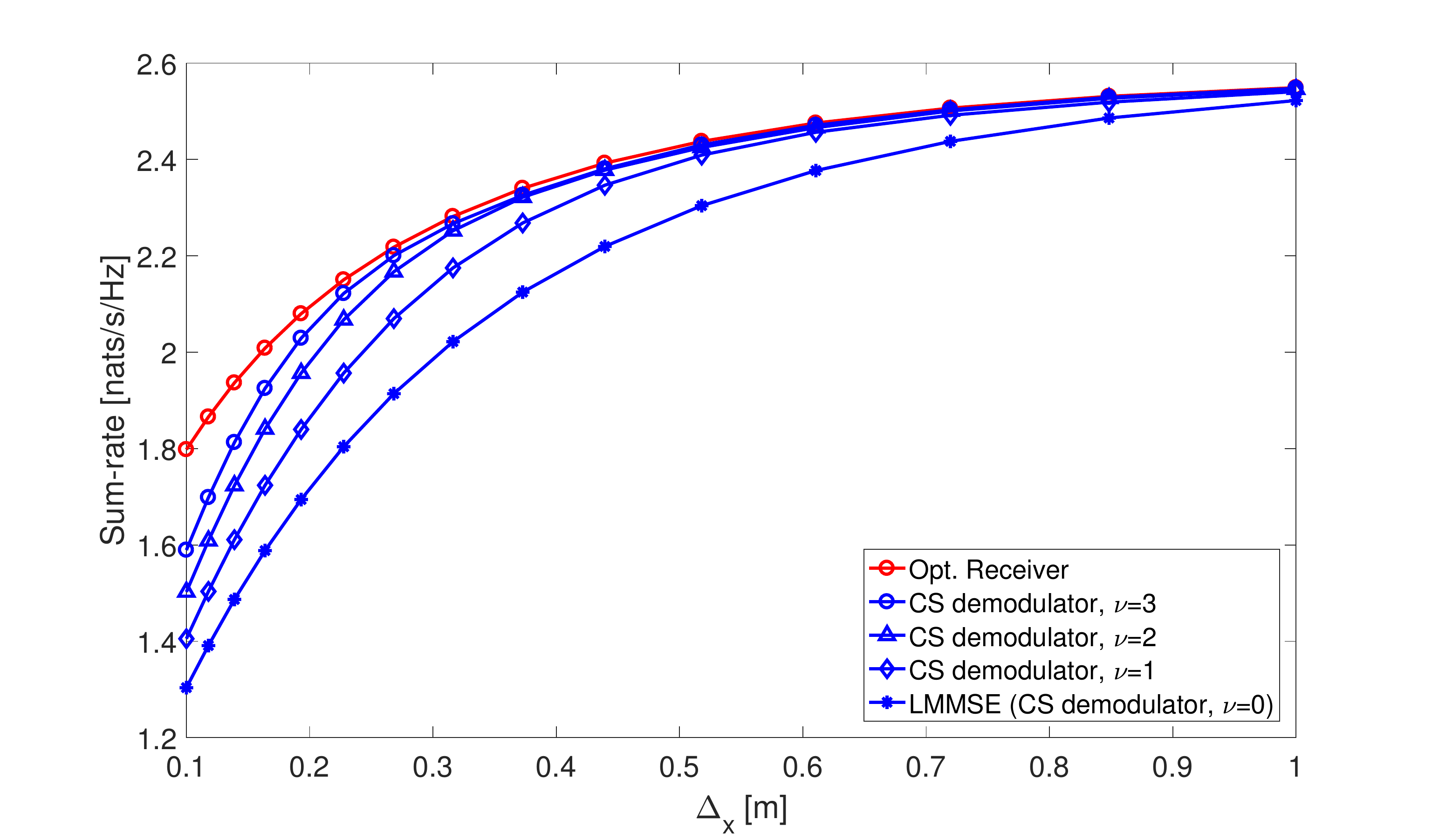}}
\vspace*{-7mm}
\caption{\label{figCS}The sum-rate with CS demodulator and a rectangular LIS of size $A\!=\!B\!=\!1$ m deployed in a room of length 4 m, width 8 m and height 4 m. We set the noise PSD $N_0\!=1$, $\!\lambda\!=\!0.5$ m and $\hat{P}\!=\!10$.}
\vspace*{-6mm}
\end{center}
\end{figure}

\begin{figure}
\begin{center}
\vspace*{-4mm}
\hspace*{-2mm}
\scalebox{.42}{\includegraphics{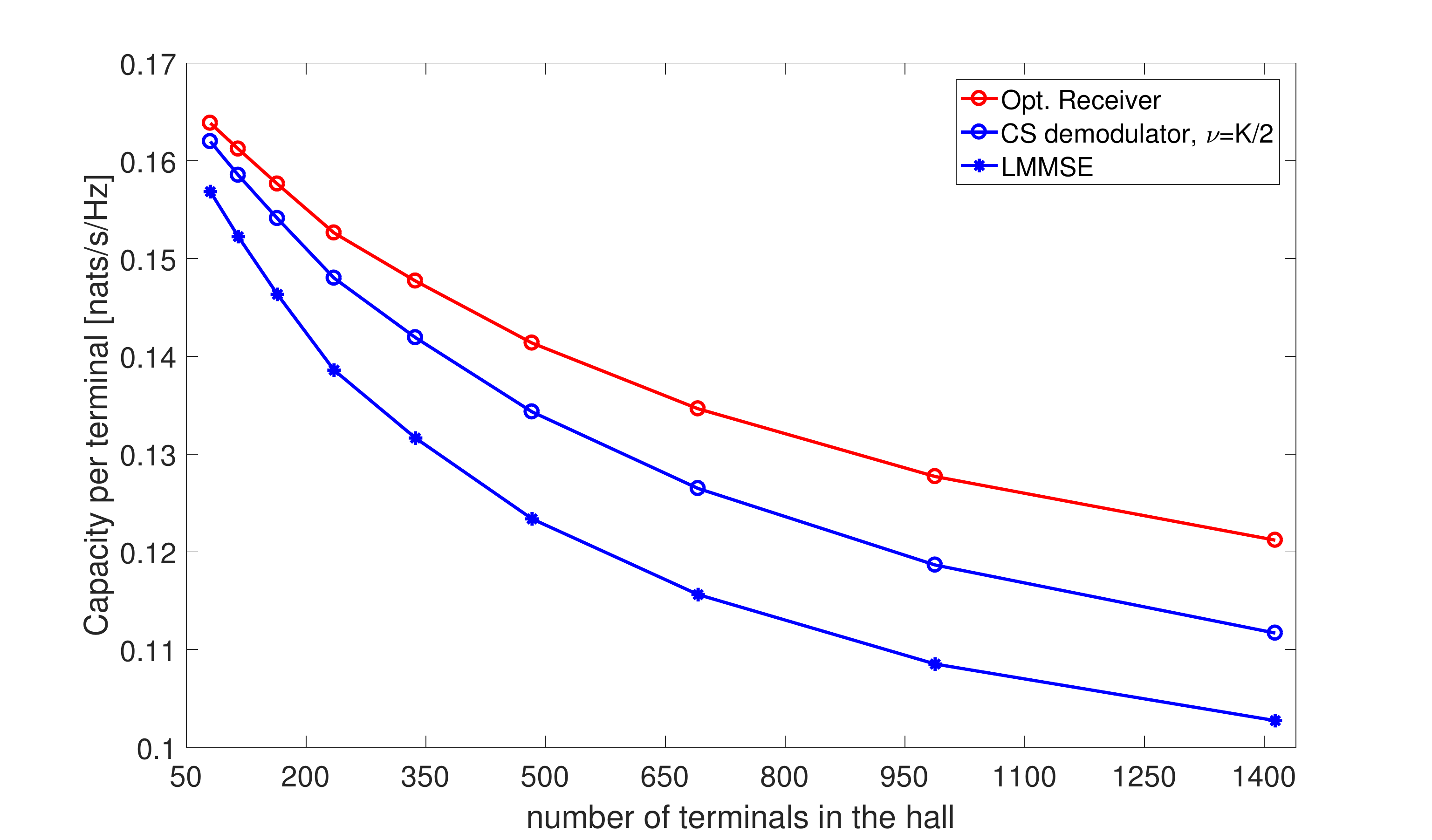}}
\vspace*{-7mm}
\caption{\label{fig10}Evaluating the capacity per terminal under the same settings as in Fig. \ref{figCS} with two-dimensional terminal-deployments and for increasing terminal-density.}
\vspace*{-12mm}
\end{center}
\end{figure}

\subsection{Capacities with CS Demodulator for LIS}
At last, we evaluate the sum-rate with a rectangular LIS of size $A\!=\!B\!=\!1$ m deployed in a room of length 8 m, width 8 m and height 4 m, and with $N_0\!=1$, $\!\lambda\!=\!0.5$ m and $\hat{P}\!=\!10$. We compare the sum-rates obtained with the optimal receiver and the CS demodulators where 15 terminals are located on the ground in a line and with spacing $\Delta_x$, while the LIS is deployed in the center of the roof. As can be seen in Fig. \ref{figCS}, when $\Delta_x$ is larger than $\lambda/2\!=\!0.25$ m, the CS demodulator with $\nu\!=\!1$ converges fast to the optimal receiver, but has a much less complexity.

In Fig. \ref{fig10}, we compare the capacities per terminal obtained with the optimal receiver and the LMMSE demodulator when terminal-density increases under the same settings as in Fig. \ref{figCS}, but with a two-dimensional terminal-deployment where terminals are uniformly located on the ground. When there are 640 terminals (that is, 10 terminals per m$^2$), the LMMSE demodulator renders approximately 14\% capacity loss compared to the optimal receiver, which is due to the limited size of the LIS. Ideally, with an infinitely large LIS, the independent signal dimensions is $\pi/\lambda^2\!=\!13$ per m$^2$ deployed surface-area as shown in Property 5. We acknowledge that, setting $\nu\!=\!K/2$ for a CS demodulator reduces the performance gaps to the optimal receiver to be less than 5\%. However, since $\nu$ is large, the CS demodulator becomes over complex unless a low modulation scheme is used. Nevertheless, the LMMSE demodulator performs reasonably well due to the high efficiency of the LIS in suppressing the interference as seen in Fig. \ref{fig9}-\ref{fig10}.

\section{Summary}

In this paper, we have considered using large intelligent surfaces (LIS) as large antenna array systems for data-transmissions with multiple single-antenna autonomous terminals. We have shown that, under the constraint that the transmit power per volume-unit $\hat{P}$ is fixed, the limit of a space-normalized capacity $\hat{\mathcal{C}}$ per volume-unit is $\hat{P}/(2N_0)$ when the wavelength $\lambda$ approaches zero. We have also derived that the numbers of independent signal dimensions can be harvested for different terminal-deployments, which are shown to be $\frac{2}{\lambda}$ per meter (m) for one-dimensional terminal-deployment, and $\pi/\lambda^2$ per m$^2$ for two and three dimensional cases. 

We have also analyzed the optimal sampling lattice for designing the LIS in a practical system based on sampling theory and shown that, the hexagonal lattice is optimal for minimizing the surface-area of a LIS under the constraint that one independent signal dimension should be obtained per spent antenna. Further, we have also extensively discussed low-complexity channel shortening (CS) demodulator design for the data-transmissions with the LIS. In addition, we shown through numerical results that the LIS provides robust performances when the number of terminals increases and is highly effective in interference suppressing, which makes it a promising direction of research for data-transmission in communication systems beyond massive-MIMO.

\appendices

\section*{Appendix A: Argumentations of Property 1}
We first define a function $\varphi$ as
\bea \label{appA1} \varphi(x) =\left(1+x^2\right)^{-\frac{3}{4}}\exp\!\left(\!-\frac{2\pi\jmath}{\lambda}\sqrt{1+x^2}\right). \eea 
\hspace{-1.4mm}Then the function $g(\Delta)$ can be written as 
\bea \label{appA2} g(\Delta) =\int_{-\infty}^{\infty}\varphi(x) \varphi^{\ast}(x+\Delta) \mathrm{d}x
= \varphi(\Delta)\star\varphi^{\ast}(\Delta). \eea
To show that $g(\Delta)$ is close to a sinc-function, we first need to show that the Fourier transform of $\varphi(x)$ is close to a brick-shape. The Fourier transform $\varPhi(f)$ is
{\setlength\arraycolsep{2pt} \bea \label{appA3} \varPhi(f) &=& \int_{-\infty}^{\infty}\varphi(x)\exp\!\left(- 2\pi \jmath fx\right)\mathrm{d}x  \notag \\
&=&2\int_{0}^{\infty}\left(1+x^2\right)^{-\frac{3}{4}}\exp\!\left(\!-\frac{2\pi\jmath}{\lambda}\sqrt{1+x^2}\right)\cos\left(2\pi fx\right)\mathrm{d}x  . \eea}
\hspace{-1.4mm}Noticing that the following Fourier cosine transforms (FCTs) hold
{\setlength\arraycolsep{2pt} \bea \label{FCTK0}\int_{0}^{\infty}\left(1+x^2\right)^{-\frac{1}{2}}\exp\!\left(\!-\frac{2\pi\jmath}{\lambda}\sqrt{1+x^2}\right)\cos\left(2\pi fx\right)\mathrm{d}x&=&K_0\left(2\pi\sqrt{f^2\!-\!\lambda^{-2}} \right)\!,  \\
\label{IFCTK0} 4\int_{0}^{\infty}K_0\left(2\pi\sqrt{f^2-\lambda^{-2}} \right)\cos\left(2\pi fx\right)\mathrm{d}f&=&\left(1+x^2\right)^{-\frac{1}{2}}\exp\!\left(\!-\frac{2\pi\jmath}{\lambda}\sqrt{1+x^2}\right)\!,  \qquad  \eea}
\hspace{-1.4mm}we then have
{\setlength\arraycolsep{2pt} \bea \label{Phi_f} \varPhi(f)
&=&2\int_{0}^{\infty}\left(1+x^2\right)^{-\frac{3}{4}}\exp\!\left(\!-\frac{2\pi\jmath}{\lambda}\sqrt{1+x^2}\right)\cos\left(2\pi fx\right)\mathrm{d}x\notag \\ &&
=8 \int_{0}^{\infty} \!\int_{0}^{\infty}\left(1+x^2\right)^{-\frac{1}{4}}K_0\left(2\pi\sqrt{\xi^2\!-\!\lambda^{-2}} \right)\cos\left(2\pi \xi x\right)\cos\left(2\pi f x\right)\mathrm{d}\xi\mathrm{d}x\notag \\
&&=4\int_{0}^{\infty}K_0\left(2\pi\sqrt{\xi^2\!-\!\lambda^{-2}} \right) \!\big(\Phi_c\big(|f\!+\!\xi|\big)+\Phi_c\big(|f\!-\!\xi|\big)\big)\mathrm{d}\xi\notag \\
&&=4\left(K_0\left(2\pi\sqrt{f^2\!-\!\lambda^{-2}} \right) \!\star\Phi_c\big(|f|\big)\right), \eea}
\hspace{-1.4mm}where $\Phi_c(f)$ is the FCT of $\left(1+x^2\right)^{-\frac{1}{4}}$, which is an even functions and for $f\!\geq \!0$ equal to
 {\setlength\arraycolsep{2pt}\bea  \label{phi_c} \Phi_c(f)&=&\int_{0}^{\infty}\left(1+x^2\right)^{-\frac{1}{4}}\cos(2\pi fx)\mathrm{d}x \notag \\
 &=& \frac{\left(\pi/f\right)^{\frac{1}{4}}}{\Gamma\big(\frac{1}{4}\big)}K_{\frac{1}{4}}(2\pi f ). \eea}
\hspace{-1.4mm}The functions $K_0(f)$and $K_{\frac{1}{4}}(f)$ are the modified Bessel function of the second kind, A closed-from expression of the convolution in (\ref{Phi_f}) seems out of reach and we have to seek an approximation of it. Firstly, noticing that the amplitude of the modified Bessel function $K_0\big(2\pi\sqrt{f^2\!-\!\lambda^{-2}}\big)$ is lower-bounded by a rectangular function as
\bea \label{absK0} \left|K_0\big(2\pi\sqrt{f^2\!-\!\lambda^{-2}}\big)\right| \approx\left\{ \begin{array}{cc}\frac{\sqrt{\lambda}}{2},&-\frac{1}{\lambda}<f<\frac{1}{\lambda} \\0,  &\mathrm{otherwise}. \end{array}  \right.  \eea

\begin{figure}[t]
\begin{center}
\vspace*{-4mm}
\hspace*{-2mm}
\scalebox{.42}{\includegraphics{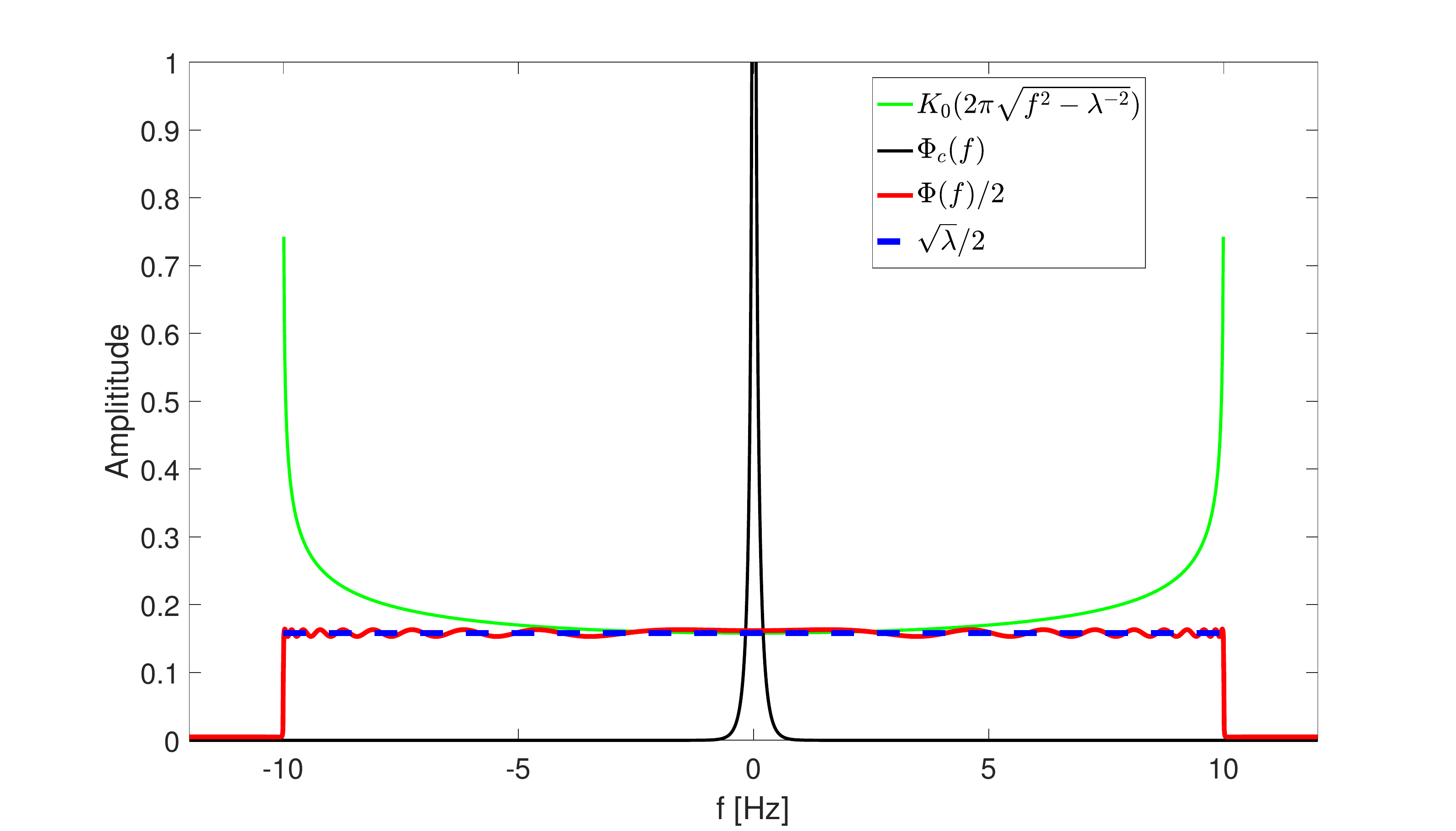}}
\vspace*{-7mm}
\caption{\label{appA}The function values of $K_0\big(2\pi\sqrt{f^2\!-\!\lambda^{-2}}\big)$ and $\Phi_c(f)$ and the sinc-function approximation of $\Phi(f)$ for $\lambda\!=\!0.1$. Note that $f\!=\!0$ is a singularity point for $\Phi_c(f)$ and $f\!=\!\pm\frac{1}{\lambda}$ are singularity points for $K_0\big(2\pi\sqrt{f^2\!-\!\lambda^{-2}}\big)$.}
\vspace*{-6mm}
\end{center}
\end{figure}

\begin{figure}
\begin{center}
\vspace*{-4mm}
\hspace*{-2mm}
\scalebox{.41}{\includegraphics{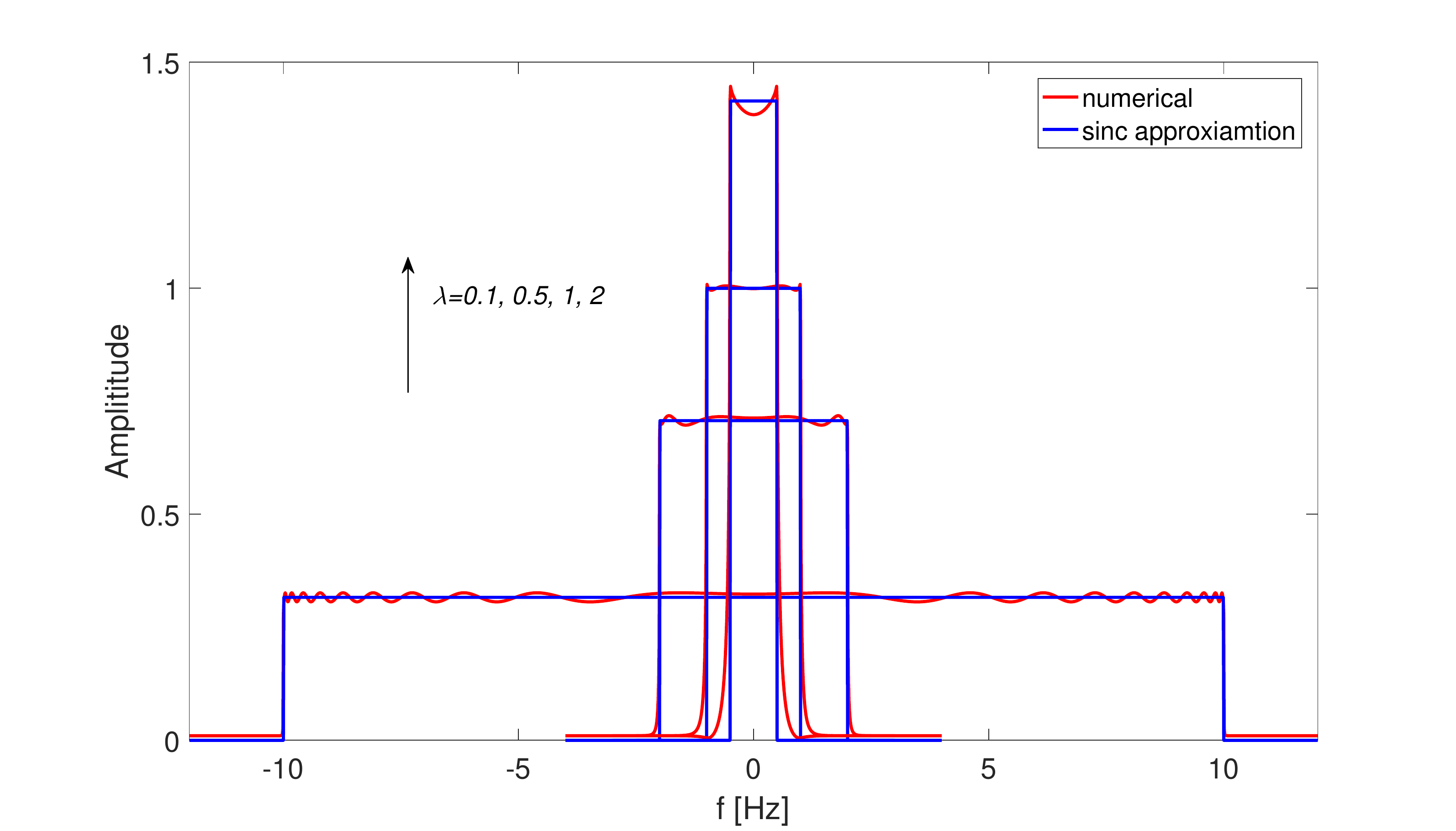}}
\vspace*{-7mm}
\caption{\label{appB}The numerical integrals and approximations with the sinc-function of $\varPhi(f)$ for different values of $\lambda$.}
\vspace*{-12mm}
\end{center}
\end{figure}

Secondly, $\Phi_c(f)$ can be approximated with a Dirac delta-function when $\lambda$ is small\footnote{In order to approximate $\Phi_c(f)$ by a Dirac delta-function in (\ref{Phi_delta}), the bandwidth of $K_0\big(2\pi\sqrt{f^2\!-\!\lambda^{-2}}\big)$ should be much larger than that of $\varPhi_c(f)$, that is to say, $\lambda$ should not be too small. As shown in the Fig. \ref{appB}, the sinc-function approximation of $\varPhi(f)$ works well with $\lambda$ up to 1 m.} as
\bea \label{Phi_delta} \Phi_c(f)\approx\frac{1}{2}\delta(f) .\eea 
Then, it holds that
 {\setlength\arraycolsep{2pt} \bea \label{Phi_approx} \varPhi(f)&\approx&2\int_{-\infty}^{\infty}K_0\big(2\pi\sqrt{\xi^2-\lambda^{-2}}\big)\delta(f+\xi)\mathrm{d}\xi \notag \\
&=&2K_0\big(2\pi\sqrt{f^2-\lambda^{-2}}\big). \qquad \eea}
\hspace{-1.4mm}From (\ref{appA2}), the Fourier transform of $g(\Delta)$, which is denoted as $G(f)$, equals
 \bea \label{Gf} G(f)=|\varPhi(f)|^2\approx\left\{ \begin{array}{cc}\lambda,&-\frac{1}{\lambda}<f<\frac{1}{\lambda} \\0,  &\mathrm{otherwise}\end{array}  \right.  \eea
which is also a rectangular function and the inverse Fourier transform is
 \bea  g(\Delta)\approx 2 \mathrm{sinc}\left(\frac{2\pi\Delta}{\lambda}\right).\notag \eea

In Fig. \ref{appB}, we plot the numerical computation of $\varPhi(f)$ and the sinc-function approximations, where the capacities given by the PSD of $|\varPhi(f)|^2$, computed as $\int\limits_{-\infty}^{\infty}\log\left(1+|\varPhi(f)|^2\right)\mathrm{d}f$, are equal to $[1.9269,\,1.6383,\, 1.4044,\,1.1334]$ nats/s/Hz for $\lambda\!=\![0.1,\,0.5,\, 1,\,2]$ m, while the capacities with sinc-function based approximations, computed as $\int\limits_{-\infty}^{\infty}\log\left(1+|\mathrm{sinc}(f)|^2\right)\mathrm{d}f$, are equal to $[1.9053 ,\,1.6178,\, 1.3794,\,1.0876]$ nats/s/Hz, which are close and slightly smaller.

\section*{Appendix B: Proof of Property 2}
From (\ref{glk}), it holds that 
 {\setlength\arraycolsep{2pt}  \bea \label{int2} g_{k,\ell}&=&P\iint\limits_{(x,\,y)\in\mathcal{S}} s^\ast_{x_k,y_k,z_k}(x,y)s_{x_\ell,y_\ell,z_\ell}(x,y)\mathrm{d}x \mathrm{d}y \notag \\
 &=&\frac{z_0P}{4\pi}\int\limits_{-B}^{B}\!\int\limits_{-\infty}^{\infty} \frac{1}{(\eta_k\eta_\ell)^{\frac{3}{4}}}\exp\!\left(\!\frac{2\pi j(\sqrt{\eta_k}-\sqrt{\eta_\ell}}{\lambda}\right)\mathrm{d}x \mathrm{d}y. \eea}
\hspace*{-1.4mm}where the metrics $\eta_k$ and $\eta_\ell$ equal
 {\setlength\arraycolsep{2pt}\bea \eta_k&=&z_0^2+y^2+(x-x_k)^2, \\
  \eta_\ell&=&z_0^2+y^2+(x-x_\ell)^2.   \eea}
\hspace*{-1.4mm}Using Property 1, it can be shown that
\bea \label{int1} \int\limits_{-\infty}^{\infty} \frac{1}{(\eta_k\eta_\ell)^{\frac{3}{4}}}\exp\!\left(\!\frac{2\pi j(\sqrt{\eta_k}-\sqrt{\eta_\ell}}{\lambda}\right)\mathrm{d}x =\frac{2}{z_0^2+y^2}\mathrm{sinc}\left(\frac{2(x_k-x_\ell)}{\lambda}\right)\!. \eea
Inserting (\ref{int1}) back into (\ref{int2}) yields
 \bea g_{k,\ell}=\frac{z_0P}{2\pi}\mathrm{sinc}\left(\frac{2(x_k-x_\ell)}{\lambda}\right)\int\limits_{-B}^{B}\frac{1}{z_0^2+y^2}\mathrm{d}y 
 =\frac{P}{\pi}\tan^{-1}\left(\frac{B}{z_0}\right)\mathrm{sinc}\left(\frac{2(x_k-x_\ell)}{\lambda}\right),    \eea
which completes the proof.

\section*{Appendix C: Proof of Property 3}
We first define another auxiliary parameter $\tilde{\theta}\!=\!1\!-\!\beta\theta\!=\!\alpha\theta$. From the definition of $G(f)$ in (\ref{Gf}), the capacity (\ref{Copt}) can be split into two parts. In a first part, $G(f)$ is folded by $\beta$ times with amplitude $\beta\theta \zeta P$ and the integration interval length being $\theta-\tilde{\theta}$, and in a second part, $G(f)$ is folded by $\beta\!+\!1$ times with amplitude $(\beta\!+\!1)\theta \zeta P$ and the integration interval length being $\tilde{\theta}$. Hence, the capacity (\ref{Copt}) equals
 \bea \mathcal{C}\!=\! \frac{1}{\theta}\Bigg(\!(\theta-\tilde{\theta})\log\!\left(\!1\!+\!\frac{\beta\theta \zeta P }{N_0}\!\right)+\tilde{\theta}\log\!\left(\!1\!+\!\frac{(\beta\!+\!1)  \theta \zeta P}{ N_0}\!\right)\!\!\Bigg)\!.  \eea
By the definition of $\alpha$, $\beta$ in (\ref{a}) and utilizing (\ref{theta}) yields the capacity stated in Property 3.

\section*{Appendix D: Proof of Property 4}
With only the MF procedure, the capacity with ISI present is in (\ref{Cmf}), where the interference can be expressed as
\bea  \label{intpow} I=\frac{1}{\zeta P}\sum\limits_{\ell=-\infty, \ell\neq 0}^{\infty}\left|g_\ell\right|^2=\frac{1}{\theta \zeta P}\int_{-\frac{\theta}{2}}^{\frac{\theta}{2}}\!\left|G(f)\right|^2\mathrm{d}f-\zeta P.\eea
The second equality in (\ref{intpow}) is from Parseval\rq{}s identity applied to $G(f)$ in (\ref{Gf}). Following the same arguments of $G(f)$ as proving Property 3, the interference power can be written as
  {\setlength\arraycolsep{2pt} \bea  \label{intpow2} I&=&\frac{1}{\theta \zeta P}\left( (\theta-\tilde{\theta})(\beta\theta \zeta P )^2+\tilde{\theta}((\beta+1)\theta \zeta P )^2\right)-\zeta P \notag \\
&=&\theta \zeta P \left( \theta\beta^2+2\tilde{\theta}\beta +\tilde{\theta}\right)-\zeta P.\eea}
\hspace{-1.4mm}As $\tilde{\theta}\!=\!\alpha\theta$, inserting it back into (\ref{intpow2}) yields the expression of $I$ in (\ref{I}).

\section*{Appendix E: Proof of Property 6}
The proof considers, without loss of generality, $\lambda=1$. A simple scaling gives the result for arbitrary $\lambda$. In two dimensions, the fundamental cell is always a centrally-symmetric hexagon (possibly degenerating into a rectangle) inscribed in a circle whose radius, is the circumcenter of a triangle with vertices $0,\vec{v},\vec{w}$ for some vectors $\vec{v},\vec{w}$ that generate the lattice. So the volume of the lattice generated from $\vec{S}^{-T}$ is twice the area of a triangle inscribed in a unit circle, and this area is maximized when the triangle is equilateral. This makes the lattice generated from $\vec{S}^{-T}$, and thus also from $\vec{S}$, hexagonal.

\end{document}